\documentclass[%
 reprint,
 amsmath,amssymb,
 aps,
]{revtex4-2}

\usepackage{graphicx}
\usepackage{dcolumn}
\usepackage{bm}
\usepackage[hidelinks]{hyperref}

\usepackage{braket}
\usepackage[version=4]{mhchem}
\usepackage[capitalize]{cleveref}
\usepackage{mathtools}
\def\equationautorefname#1#2\null{%
  Eq.#1(#2\null)%
}
\def\tableautorefname#1#2\null{%
  {table#1#2\null}%
}
\def\sectionautorefname#1#2\null{%
  {Sec.#1#2\null}%
}
\def\figureautorefname#1#2\null{%
  {Fig.#1#2\null}%
}
\def\subsectionautorefname#1#2\null{%
  {Subsec.#1#2\null}%
}

\begin{document}

\preprint{APS/Finite_Resolution}

\title{Entropy estimation for partially accessible Markov networks based on imperfect observations: Role of finite resolution and finite statistics}

\author{Jonas H. Fritz}
\author{Benjamin Ertel}%
\author{Udo Seifert}%
\affiliation{%
 II. Institut für Theoretische Physik, Universität Stuttgart, 70550 Stuttgart, Germany
}%

\date{\today}

\begin{abstract}
Estimating entropy production from real observation data can be difficult due to finite resolution in both space and time and finite measurement statistics. We characterize the statistical error introduced by finite sample size and compare the performance of three different entropy estimators under these limitations for two different paradigmatic systems, a four-state Markov network and an augmented Michaelis-Menten reaction scheme. We consider the thermodynamic uncertainty relation, a waiting-time based estimator for resolved transitions and a waiting-time based estimator for blurred transitions in imperfect observation scenarios. For perfect measurement statistics and finite temporal resolution, the estimator based on resolved transitions performs best in all considered scenarios. The thermodynamic uncertainty relation gives a better estimate than the estimator based on blurred transitions at low driving affinities, whereas the latter performs better at high driving affinities. Furthermore, we find that a higher temporal and spatial resolution leads to slower convergence of measurement statistics, implying that for short measurement times, a lower resolution may be beneficial. Additionally, we identify a self-averaging effect for the waiting-time based entropy estimators that can reduce their variance for observations with finite statistics.
\end{abstract}

\maketitle

\section{Introduction}

Within the framework of stochastic thermodynamics \cite{Sekimoto2010,jarzynski2011,seifert2012,broeck2015}, the entropy production can be quantified based on a Markovian description, which leads to applications for various chemical and biological systems, including chemical reaction networks \cite{schmiedl2007,ge2012,rao2016} and molecular motors \cite{qian2000,andrieux2006,seifert2011,chowdhury2013,kolomeisky_book,speck2021}. In practice, evaluating the entropy production can be challenging because the Markovian description might only be partially accessible. For instance, in experimental studies of molecular motors, the motion of the motor is registered through an attached bead, while the concurrent chemical reactions are not resolved \cite{ritort2006,herbert2008,veigel2011,ariga2018,bustamante2022}.

For inferring the entropy production in these partially accessible systems, numerous estimators emphasizing different observables have been proposed. State-lumping estimators \cite{mehl2012,bo2014,bo2017,uhl2018,lucente2022,esposito2012} bound the entropy production from below based on the apparent entropy production of an effective Markovian description \cite{rahav2007,gomez-marin2008,esposito2012}. If invasive manipulations are applied, this bound can be tightened \cite{polettini2017,bisker2017}. Decimation schemes and optimization techniques \cite{pigolotti2008,puglisi2010,teza2020,skinner2021_1,ehrich2021,nitzan2023} can increase the quality of these estimators as well. Other estimators based on the same type of effective description emerge from fluctuation theorems for time-antisymmetric observables \cite{kawaguchi2013,shiraishi2015}. The thermodynamic uncertainty relation (TUR) establishes a fundamental lower bound for the entropy production in terms of fluctuation statistics of observable currents \cite{barato2015,gingrich2016,pietzonka2016,horowitz2020}. Statistical quantities and general counting observables \cite{roldan2010,roldan2012,biddle2020,pietzonka2021,pietzonka2023,Diterlizzi2024,Diterlizzi2024_2}, extreme value statistics \cite{neri2017,neri2020,neri2022,polettini2024} and correlations \cite{oberreiter2022,ohga2023,dechant2023} yield entropy estimators as well. Incorporating waiting-times into effective descriptions introduces a complementary approach to entropy estimation \cite{berezhkovskii2019,martinez2019,skinner2021_2,hartich2021_2,hartich2021}. Effective descriptions with renewal processes \cite{qian2007,esposito2008,maes2009,PRE} allow for inference, provided observables can be identified as Markovian events based on their time-reversal operation \cite{hartich2021_comment,bisker2022_Reply,PRL,PRR,PNAS}. In particular, descriptions based on the observation of waiting-times between transitions lead to a lower bound for the entropy production \cite{PRX,harunari2022}. Moreover, these types of waiting-times yield additional topological information about the underlying system \cite{PRX,cao2008,li2013,panigrahy2019,thorneywork2020,berezhkovskii2020,satija2020,IJMS}. 

Conceptually, all of these entropy estimators rely on the assumption of ideal statistics, i.e., they consider observations with arbitrary precision. In reality, however, the accuracy of measurements is limited. As an example, consider the observation of enzyme dynamics in the context of single-molecule enzyme kinetics. While theoretical results typically assume a Markov network as description, experimental observations often yield bulk quantities with finite temporal and spatial resolution \cite{fersht2002,Illanes2008,cornish2012}. Even if single molecules can be resolved, applying these estimation techniques is not straightforward, see Ref. \cite{godec2022}. For imperfect observations, specialized entropy estimators have been developed to handle specific types of limited precision \cite{baiesi2024,blom2024,song2024}. This class of estimators includes an estimator for the observation of blurred transitions in Markov networks, i.e., for transitions which are not distinguishable due to finite resolution in the space of observable transitions \cite{PRE2,harunari2024}.

This work aims at illustrating and understanding the impact of imperfect observations on extant entropy estimators, rather than introducing a new estimator or bounding errors as done in Ref. \cite{bebon2023}. We consider observation scenarios with finite temporal and spatial resolution as well as finite observation statistics for two different systems. The first system, a four-state Markov network, illustrates the main concepts. The second system, a reaction network based on the Michaelis-Menten mechanism serves as paradigmatic example for more realistic observations. The paper is structured as follows. In \autoref{SEC:Setup}, we recapitulate concepts from stochastic thermodynamics and introduce the entropy estimators we consider. Furthermore, we discuss the impact of limited temporal and spatial resolution. We extend this discussion in \autoref{SEC:FStat} by considering finite statistics. In \autoref{SEC:Markov} and \autoref{SEC:Micha}, we present and interpret simulation results for the two systems. The final \autoref{SEC:Conclusion} contains a concluding perspective.

\section{Setup}
\label{SEC:Setup}
\subsection{Stochastic thermodynamics}
\begin{figure}[bt]
    \includegraphics[width=0.5\linewidth]{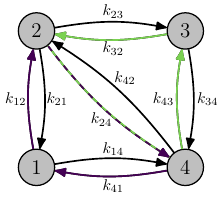}
    \caption[Setup figure 1]{Four-state Markov network with ten possible transitions characterized by transition rates $k_{ij}$ along its five different edges. The network includes three different cycles $\mathcal{C}_1 = 1241$ (dark color), $\mathcal{C}_2 = 3243$ (bright color) and $\mathcal{C}_3 = 12341$. The affinities $\mathcal{A}_{\mathcal{C}_1}$ and $\mathcal{A}_{\mathcal{C}_2}$ determine the value of $\mathcal{A}_{\mathcal{C}_3}$ uniquely due to the topology of the system.}
\label{Fig:SETUP_Markov}
\end{figure}

We consider a Markov network with $N$ states for which a transition $(ij)$ between state $i$ and state $j$ happens with rate $k_{ij}$ along the corresponding edge or, equivalently, link of the network. For thermodynamic consistency, we assume that each rate $k_{ij} > 0$ implies $k_{ji} > 0$ for the reversed transition $(ji)$. A Markov network which illustrates these quantities is shown in \autoref{Fig:SETUP_Markov}. If all rates are time-independent, the dynamics will eventually reach a stationary state in the long-time limit $t\to\infty$ which becomes a non-equilibrium stationary state (NESS) if detailed balance is broken for at least one edge. A NESS is characterized by a mean entropy production rate \cite{seifert2012}
\begin{equation}
    \braket{\sigma} = \sum_{(ij)}p_i^sk_{ij}\ln\frac{p_i^sk_{ij}}{p_j^sk_{ji}} = \sum_{(ij)}\pi_{ij}\ln\frac{\pi_{ij}}{\pi_{ji}}\geq 0,
    \label{EQ:SETUP_Ep1}
\end{equation}
where $p_i^s$ is the stationary distribution of state $i$ and the second equality defines the stationary rate of observation $\pi_{ij} \equiv p_i^sk_{ij}$ for transition $(ij)$. These rates set the timescale $\langle\tau\rangle$ of transitions in the network via $1/\langle\tau\rangle\equiv\sum_{(ij)}\pi_{ij}$. Note that here and in the following, entropy is dimensionless and we set the inverse thermal energy to $\beta = 1$. We denote averages over many realizations of the system with $\braket{...}$. 

In the NESS, the entropy production is caused by cycle currents within the Markov network. Each cycle current $j_{\mathcal{C}}$ originates from a cycle $\mathcal{C}$, which is a closed and directed loop without self-crossings. The network in \autoref{Fig:SETUP_Markov} includes three different cycles. By favoring the forward direction over the reverse or vice versa, these cycles can break detailed balance and thus time-reversal symmetry. This preference for one direction is quantified by a non-vanishing cycle affinity
\begin{equation}
    \mathcal{A}_{\mathcal{C}} = \prod_{(ij)\in\mathcal{C}}\ln\frac{k_{ij}}{k_{ji}},
    \label{EQ:SETUP_Aff}
\end{equation}
where the numerator includes all forward rates of the cycle and the denominator includes the corresponding backward rates. By summing the contributions from all cycles, $\braket{\sigma}$ can be expressed as
\begin{equation}
    \braket{\sigma} = \sum_{\mathcal{C}}j_{\mathcal{C}} \mathcal{A}_{\mathcal{C}}.
    \label{EQ:SETUP_Ep2}
\end{equation}
Operationally, the current of each cycle is given by the mean net number of cycle completions divided by the observation time \cite{hill1989,jiang2004}. The affinity of each cycle can be interpreted as the entropy production associated with a single completion of the cycle \cite{schnakenberg1976,hill1989}. In chemical reaction networks, for example, $j_{\mathcal{C}}$ tracks the net creation and annihilation of molecules during each reaction cycle, while $\mathcal{A}_{\mathcal{C}}$ quantifies the corresponding change in the free energy of the system \cite{schmiedl2007,rao2016,seifert2012}.

\begin{figure*}[bt]
    \centering
    \includegraphics[width=0.75\linewidth]{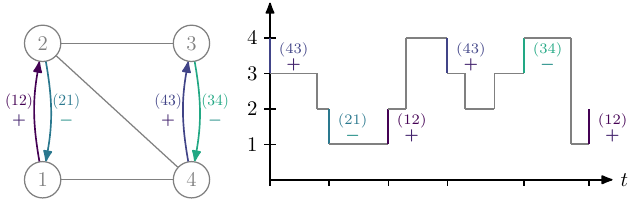}
    \caption[Setup figure 2]{Observation of the transitions $(12),(21),(34)$ and $(43)$ in the four-state Markov network from \autoref{Fig:SETUP_Markov}. In the course of time, this observation leads to a trajectory consisting of observed transitions $(ij)$ and in-between waiting-times $t_i$. Counting $(12)$ and $(43)$ as $+$-transitions and $(21)$ and $(34)$ as $-$-transitions along this trajectory results in a generalized current $\mathcal{J} = n_+ - n_-$ counting cycle completions. Incorporating the waiting-times between observed transitions leads to 16 different waiting-time distributions $\psi_{(ij)\to (kl)}(t)$ for observed transition sequences $(ij)\to (kl)$.}
    \label{Fig:SETUP_Estimation}
\end{figure*}
\subsection{Entropy estimation}
Determining the mean entropy production either via \autoref{EQ:SETUP_Ep1} or via \autoref{EQ:SETUP_Ep2} requires either an observation of the complete Markov network or knowledge about all cycle currents and the corresponding affinities. If this knowledge is not accessible for a realistic system, estimation techniques must be applied. Each of these techniques provides an entropy estimator $\braket{\hat{\sigma}}$ that establishes a lower bound
\begin{equation}
    \braket{\hat{\sigma}} \leq \braket{\sigma}
    \label{EQ:SETUP_Ep3}
\end{equation}
for the entropy production of the system. In contrast to $\braket{\sigma}$, these entropy estimators rely exclusively on observable statistics and are therefore operationally accessible. We consider two different ones for Markov networks in which only a subset of the transitions is observable. To introduce these estimators through an example, we assume for the Markov network in \autoref{Fig:SETUP_Markov} that only the transitions $(12), (21), (34)$ and $(43)$ are registered by an external observer, as illustrated in \autoref{Fig:SETUP_Estimation}.

With the cycle decomposition of the entropy production from \autoref{EQ:SETUP_Ep2} in mind, the observer can interpret the registered transitions along a trajectory in terms of a generalized current 
\begin{equation}
    \mathcal{J} = \sum_{ij}n_{i\to j}q_{ij},
    \label{EQ:SETUP_J1}
\end{equation}
where $n_{i\to j}$ is the total number of jumps from $i$ to $j$ along the trajectory and $q_{ij}$ is an antisymmetric matrix of increments. For example, as illustrated in \autoref{Fig:SETUP_Estimation}, if the observer counts the transitions with respect to cycle completions, $q_{ij} = 1$ for $(12)$ and $(43)$, $q_{ij} = -1$ for $(21)$ and $(34)$ and $q_{ij} = 0$ for all other $i$ and $j$. The mean value $\braket{\mathcal{J}}$ and the variance $\text{Var}\mathcal{J}$ of any generalized current $\mathcal{J}$ establish a lower bound for the entropy production $\braket{\sigma}$ providing the entropy estimator
\begin{equation}
    \braket{\sigma_{\text{TUR}}}\equiv \frac{2\braket{\mathcal{J}}^2}{\text{Var}\mathcal{J}}\leq \braket{\sigma}.
    \label{EQ:SETUP_TUR2}
\end{equation}
This bound is one formulation of the TUR, which constrains the entropy production of a vast class of systems based on the fluctuations of generalized currents \cite{barato2015,gingrich2016,pietzonka2016,horowitz2020}.

After registering a transition $(ij)$, the external observer registers a subsequent transition $(kl)$ after time $t$. The distribution of these waiting-times is given by
\begin{equation}
    \psi_{(ij)\to (kl)}(t) \equiv p[(kl);T_{(kl)} - T_{(ij)}=t|(ij)]
    \label{EQ:SETUP_WTD}
\end{equation}
where $T_{(ij)}$ is the time at which transition $(ij)$ is registered. $\psi_{(ij)\to (kl)}(t)$ is the probability density for observing transition $(kl)$ at $T_{(kl)}=T_{(ij)}+t$ after waiting-time $t$ given transition $(ij)$ was observed at $T_{(ij)}$. These waiting-time distributions can be determined from histograms for registered transitions along trajectories. If the underlying Markov network is known, they can be calculated with the absorbing master-equation approach \cite{sekimoto2021,PRX,harunari2022}. Intuitively, the waiting-times defined in \autoref{EQ:SETUP_WTD} provide information about the underlying system, as high transition rates along short paths lead to short waiting-times while long paths with low transition rates result in long waiting-times. Based on this observation, the entropy estimator 
\begin{eqnarray}
    \braket{\sigma_{\text{WTD}}}&\equiv &\sum_{ij,kl}\int_0^{\infty}dt\pi_{ij}\psi_{(ij)\to (kl)}(t)\ln\frac{\psi_{(ij)\to (kl)}(t)}{\psi_{(\widetilde{kl})\to (\widetilde{ij})}(t)}\nonumber \\ &\leq& \braket{\sigma}
    \label{EQ:SETUP_WTD2}
\end{eqnarray}
can be derived \cite{PRX,harunari2022}. In \autoref{EQ:SETUP_WTD2}, the summation includes all observed transitions and $(\widetilde{ij}) = (ji)$ is the time-reversed transition of $(ij)$. If removing the observed transitions from the considered Markov network leads to a network without cycles, this estimator recovers the full entropy production, i.e., $\braket{\sigma_{\text{WTD}}} = \braket{\sigma}$ \cite{PRX,harunari2022}.

Note that although both $\braket{\sigma_{\text{TUR}}}$ and $\braket{\sigma_{\text{WTD}}}$ are estimators for the entropy production of a partially observed Markov network, their conceptual foundations are fundamentally different. From a mathematical point of view, the TUR can be derived from the Cramér-Rao inequality, a bound in estimation theory \cite{dechant2018,hasegawa2019,dechant2020}. In contrast, the waiting-time based bound relies on an effective transition-based description of the system through an appropriate renewal process with modified time-reversal operation \cite{hartich2021_2,PRX,PRL}.

\subsection{Finite resolution}
\begin{figure}[bt]
    \includegraphics[width=1.0\linewidth]{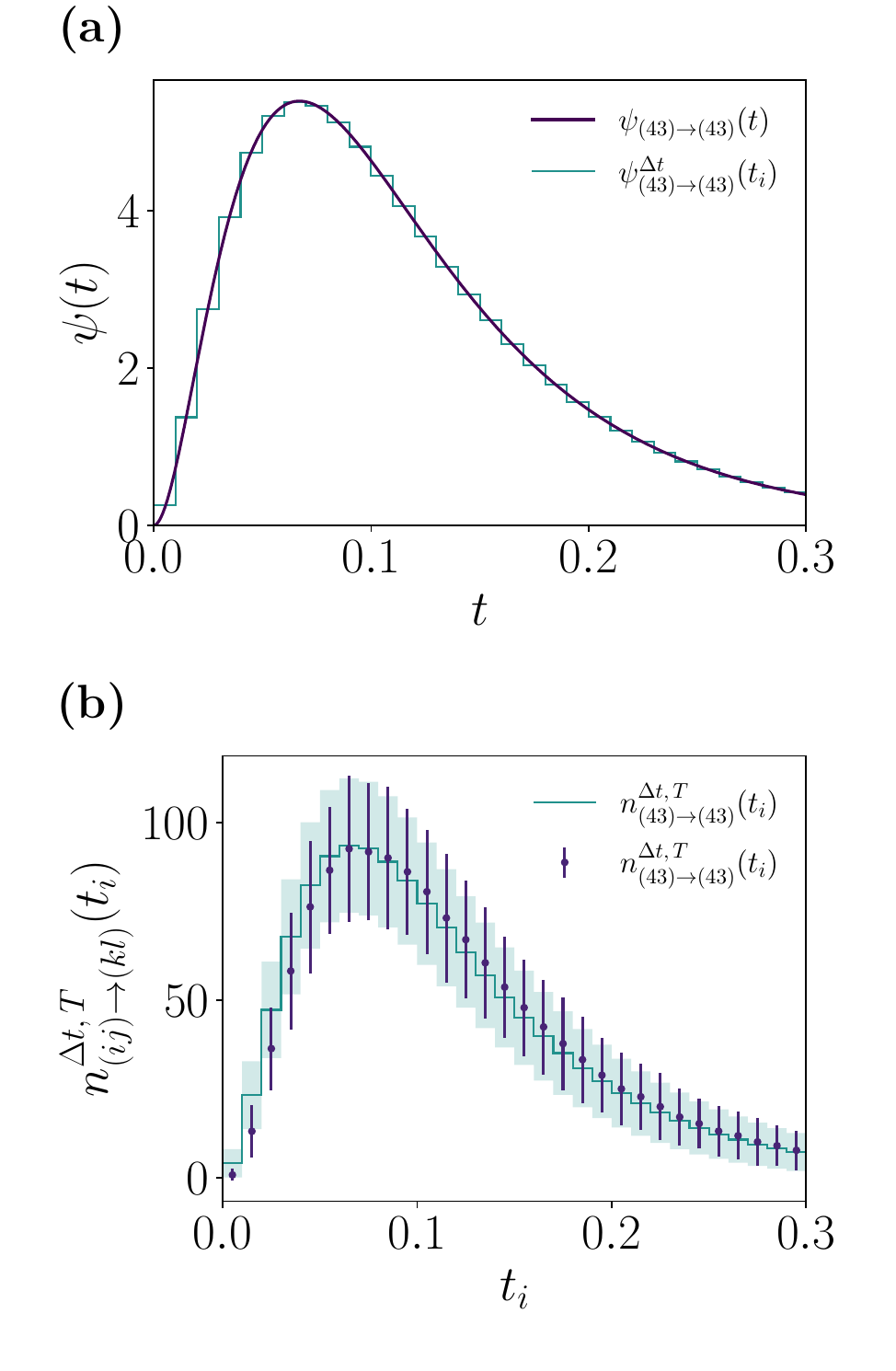}
    \caption[Setup figure 3]{Modified waiting-time distributions for the observation in \autoref{Fig:SETUP_Estimation}. (a) Approximation of waiting-time distribution $\psi_{(43)\to (43)}(t)$ at finite temporal resolution. With finite temporal resolution $\Delta t =0.01$, the waiting-time distribution (dark color) is approximated by an appropriate histogram $\psi_{(43)\to (43)}^{\Delta t}(t_i)$ (bright color). (b) Fluctuations of the histogram $n_{(43)\to (43)}^{\Delta t,\,T}(t_i)$ at finite sample size $T=300$. The analytical prediction (bright color) and a corresponding empirical histogram for $N=300$ different samples (dark color). The shaded area and error bars represent the $95\%$ confidence interval. The specific parameters are $\kappa_1=10,\,\kappa_2=1$ and $A_{\mathcal{C}_1}=7.5, \, A_{\mathcal{C}_2}=-4.5$ for the parametrization given in \cref{SEC:Markov}.}
\label{Fig:SETUP_TempCG}
\end{figure}

The entropy estimators $\braket{\sigma_{\text{TUR}}}$ and $\braket{\sigma_{\text{WTD}}}$ require accurate statistics for the mean and variance of a generalized current $\mathcal{J}$ or for the waiting-time distributions $\psi_{(ij)\to (kl)}(t)$, respectively. However, due to finite resolution, statistical accuracy in data can be limited. For example, suppose that the time difference between two consecutive visible transitions can only be measured with a finite time-resolution $\Delta t$. In this coarse-graining, we assume that an observer with finite temporal resolution still registers the true sequence of transitions, even if their time difference is smaller than $\Delta t$. Because all events are still registered, the number of transitions $n_{i\rightarrow j}$ stays the same. Thus, this type of observation yields the same generalized current $\mathcal{J}$ compared to an observation with full temporal resolution. If, in contrast, only a single transition could be registered during a time interval $\Delta t$, transitions may not be registered if two or more occur in the same measurement interval. In the resulting discrete-time Markov chain model, it is known that $\braket{\sigma_{\text{TUR}}}$ would then no longer be a lower bound for $\langle\sigma\rangle$ \cite{dechant2020}.
Therefore, we focus on the case in which the sequence and number of transitions in a measurement interval are always accessible. Then, $\braket{\sigma_{\text{TUR}}}$ is unaffected by this specific type of finite temporal resolution. In contrast, for waiting-time distributions, the finite temporal resolution implies that distributions $\psi_{(ij)\to (kl)}(t)$ need to be approximated by appropriate histograms. These histograms with lower resolution emerge from the waiting-time distributions via
\begin{equation}
    \psi_{(ij)\to (kl)}^{\Delta t}(t_i)\equiv\frac{1}{\Delta t}\int_{t_i-\Delta t/2}^{t_i+\Delta t/2}dt\,\psi_{(ij)\to (kl)}(t),
    \label{EQ:SETUP_FiniteT1}
\end{equation}
with the discrete observation times $t_i =t_0 + i\Delta t$. This approximation is illustrated in \autoref{Fig:SETUP_TempCG} (a) for $\psi_{(32)\to (32)}(t)$ from \autoref{Fig:SETUP_Estimation} using $\Delta t = 0.01$. Due to the log-sum inequality \cite{cover2006}, the inequality
\begin{eqnarray}
    \braket{\sigma_{\text{WTD}}^{\Delta t}} &\equiv &\Delta t \sum_{ij,kl,t_i}\pi_{ij} \psi_{(ij)\to (kl)}^{\Delta t}(t_i) \ln\frac{\psi_{(ij)\to (kl)}^{\Delta t}(t_i)}{\psi_{(\widetilde{kl})\to (\widetilde{ij})}^{\Delta t}(t_i)}\nonumber \\ &\leq &\braket{\sigma_{\text{WTD}}}
    \label{EQ:SETUP_FiniteT2}
\end{eqnarray}
holds, which establishes $\braket{\sigma_{\text{WTD}}^{\Delta t}}$ as an entropy estimator for $\braket{\sigma}$ based on the observation of waiting-time distributions with finite temporal resolution. In \autoref{EQ:SETUP_FiniteT2} $\pi_{ij}$ remains unchanged because this quantity does not depend on the time of observation.

Even with perfect temporal resolution, experimental limitations in an observation may still lead to finite spatial resolution. A paradigmatic example for this scenario is the observation of blurred transitions \cite{PRE2,harunari2024} where an observer is unable to distinguish between different registered transitions in a Markov network. Conceptually, the observation of blurred transitions corresponds to lumping transitions $(ij),(kl),...$ into transition classes $I,J,...$. Depending on the included transitions, these transition classes can be even or odd under time-reversal. A possible lumping for the observation in \autoref{Fig:SETUP_Estimation} is given by the observation of cycle completions in forward or backward direction as already considered for the TUR. Instead of registering single transitions $(12),(21),(34)$ or $(43)$, an observer registers transitions from the odd transition classes $+=\{(12),(43)\}$ and $-=\{(21),(34)\}$. The waiting-time distribution for the subsequent observation $I\to J$ of transition classes $I$ and $J$ is defined via \cite{PRE2}
\begin{equation}
    \Psi_{I\to J}(t)\equiv\sum_{\mathclap{\substack{(ij)\in I\\(kl)\in J}}}\frac{\pi_{ij}}{\pi_I}\psi_{(ij)\to (kl)}(t),
    \label{EQ:SETUP_FiniteT4}
\end{equation}
where $\pi_I=\sum_{ij\in I}\pi_{ij}$ corresponds to the rate of observation for transition class $I$. Although the mathematical interpretation of \autoref{EQ:SETUP_FiniteT4} is similar to \autoref{EQ:SETUP_WTD}, a lower bound in analogy to \autoref{EQ:SETUP_WTD2} does not hold. Instead, as proven in Ref. \cite{PRE2}, the two quantities
\begin{equation}
    \braket{\hat{\sigma}_{\mathrm{BT}}} \equiv \sum_{I, J}\int_0^{\infty}dt \pi_{I}\Psi_{I\to J}(t)\ln\frac{\Psi_{I\to J}(t)}{\Psi_{\widetilde{J}\to \widetilde{I}}(t)}
    \label{EQ:SETUP_FiniteT5}
\end{equation}
and
\begin{equation}
    \braket{\hat{\sigma}_{\mathrm{TC}}} = \sum_{I, J}\pi_I p_{I\to J}\log\frac{\pi_I}{\pi_{\widetilde{J}}}
    \label{EQ:SETUP_FiniteT6}
\end{equation}
provide the lower bound
\begin{equation}
    \Sigma \equiv \frac{\braket{\hat{\sigma}_{\mathrm{BT}}} + \braket{\hat{\sigma}_{\mathrm{TC}}}}{2}\leq \braket{\sigma}
    \label{EQ:SETUP_FiniteT7}
\end{equation}
for the entropy production $\braket{\sigma}$ through the observation of blurred transitions, which establishes $\Sigma$ as an estimator for the entropy production. In \autoref{EQ:SETUP_FiniteT5} and \autoref{EQ:SETUP_FiniteT6}, $\widetilde{I}$ is the time-reversed transition class of $I$ and in \autoref{EQ:SETUP_FiniteT6}, 
\begin{equation}
    p_{I\to J}\equiv\int_0^{\infty}dt \Psi_{I\to J}(t)
    \label{EQ:SETUP_FiniteT8}
\end{equation}
corresponds to the probability for observing $I\to J$ irrespective of the in-between waiting-time. With an additional temporal resolution $\Delta t$, incorporating the corresponding waiting-time distributions and transition probabilities in \autoref{EQ:SETUP_FiniteT7} leads to an entropy estimator $\Sigma^{\Delta t}$, analogous to $\braket{\sigma_{\text{WTD}}^{\Delta t}}$. In contrast to $\braket{\sigma_{\text{WTD}}}$, $\braket{\sigma_{\text{TUR}}}$ does not need to be modified for the observation of blurred transitions because by construction, generalized currents can be identified on the level of blurred transitions without limitations.

\section{Finite statistics}
\label{SEC:FStat}

A finite measurement time $T$ introduces further limitations for an observation. For this scenario, we need to consider the fluctuating, empirical quantities $\pi_{ij}^{T}$ and $\psi_{(ij)\to (kl)}^{\Delta t,\,T}(t_i)$, obtained after observing a single long trajectory of the system of length $T$. This trajectory contains a sequence of events $(ij)\overset{t_i}{\to}(kl)$ of observed transitions. Conceptually, each of these events occurs with probability $\braket{\tau }\pi_{ij}\Delta t\psi_{(ij)\to (kl)}^{\Delta t}(t_i)$. Because every transition is a renewal event, each occurrence of $(ij)\overset{t_i}{\to}(kl)$ is independent and identically distributed. Thus, the number of events $n_{(ij)\to (kl)}^{\Delta t,\,T}(t_i)$ registered in each bin of $\psi_{(ij)\to (kl)}^{\Delta t,\,T}(t_i)$ follows a binomial, i.e., in good approximation normal distribution
\begin{eqnarray}
    p(n_{(ij)\to (kl)}^{\Delta t,\,T}(t_i)) =&\phantom{}&\, \mathcal{N}(\langle n\rangle, \langle \Delta n^2\rangle)\,\cr
    \langle n \rangle=&\phantom{}&\,T \pi_{ij}\Delta t\psi_{(ij)\to (kl)}^{\Delta t}(t_i)\,\cr
    \langle \Delta n^2\rangle =&\phantom{}&\,T \pi_{ij}\Delta t\psi_{(ij)\to (kl)}^{\Delta t}(t_i)\times\cr
    &\phantom{}&\,\left[1-\langle\tau\rangle\pi_{ij}\Delta t\psi_{(ij)\to (kl)}^{\Delta t}(t_i)\right]\,.\cr
    &\phantom{}& 
\end{eqnarray}
Therefore, $n_{(ij)\to (kl)}^{\Delta t,\,T}(t_i)$ fluctuates between the observation of different trajectories. We illustrate these fluctuations of the waiting-time distribution in \cref{Fig:SETUP_TempCG} (b), where we find excellent agreement between this analytical prediction and an empirical sampling. 

Since the waiting-time distributions are fluctuating quantities, so are all derived quantities such as the entropy estimators based on these waiting-time distributions. We denote the estimator based on resolved transitions determined after observing a finite trajectory of length $T$ by
\begin{equation}
    \sigma_{\text{WTD}}^{\Delta t,\,T}\equiv \sum_{(ij),(kl),t_i}\Delta t\,\pi_{ij}^{T}\psi_{(ij)\to (kl)}^{\Delta t,\,T}(t_i)\ln\frac{\psi_{(ij)\to (kl)}^{\Delta t,\,T}(t_i)}{\psi_{(\widetilde{kl})\to (\widetilde{ij})}^{\Delta t,\,T}(t_i)},
    \label{EQ:SETUP_FiniteT3}
\end{equation}
where the missing $\braket{...}$ indicates that we no longer average over the full NESS distribution. To characterize the fluctuations of these now stochastic quantities, we consider $N$ many realizations with trajectories of length $T$. Denoting the different realizations by $\nu=1,\ldots N$, we can average according to 
\begin{equation}
    \left\langle\sigma_{\text{WTD}}^{\Delta t,\,T}\right\rangle_{N}\equiv\sum_{\nu=1}^{N}\left(\sigma_{\text{WTD}}^{\Delta t,\,T}\right)_\nu\,,
    \label{EQ:Mean_N}
\end{equation}
and calculate the corresponding variance via
\begin{equation}
    \mathrm{Var}\left( \sigma_{\text{WTD}}^{\Delta t,\,T}\right)_{N} \equiv\sum_{\nu=1}^{N} \left[\left(\sigma_{\text{WTD}}^{\Delta t,\,T}\right)_\nu-\left\langle\sigma_{\text{WTD}}^{\Delta t,\,T}\right\rangle_{N}\right]^2\,.
    \label{EQ:Var_N}
\end{equation}
As $T\to\infty$, the fluctuating quantities converge to their asymptotic values. This convergence can also be achieved in the limit $N\to\infty$, if the fixed observation time $T$ is longer than the largest timescale of observed waiting-times. Crucially, for finite observation times $T$ and finite $N$, fluctuations can lead to cases where $\left\langle\sigma_{\text{WTD}}^{\Delta t,\,T}\right\rangle_{N}\nleq \braket{\sigma_{\text{WTD}}}$ and thus $\left\langle\sigma_{\text{WTD}}^{\Delta t,\,T}\right\rangle_{N}\nleq \braket{\sigma}$ in general.

The estimator for blurred transitions is similarly affected by finite $T$ and finite $N$. More precisely, empirical values $\hat{\sigma}_{\text{BT}}^{\Delta t,\,T}$ and $\hat{\sigma}_{\text{TC}}^{\Delta t,\,T}$ with empirical waiting-time distributions need to be considered for finite statistics. Based on these quantities, we introduce the empirical, fluctuating $\Sigma^{\Delta t,\,T}$ with mean value $\left\langle\Sigma^{\Delta t,\,T}\right\rangle_{N}$ and variance $\mathrm{Var}\left( \Sigma^{\Delta t,\,T}\right)_{N}$ defined in analogy to \autoref{EQ:Mean_N} and \autoref{EQ:Var_N}, respectively.

\section{Four-state Markov network}
\label{SEC:Markov}

\begin{figure*}[bt]
    \includegraphics[width=0.95\textwidth]{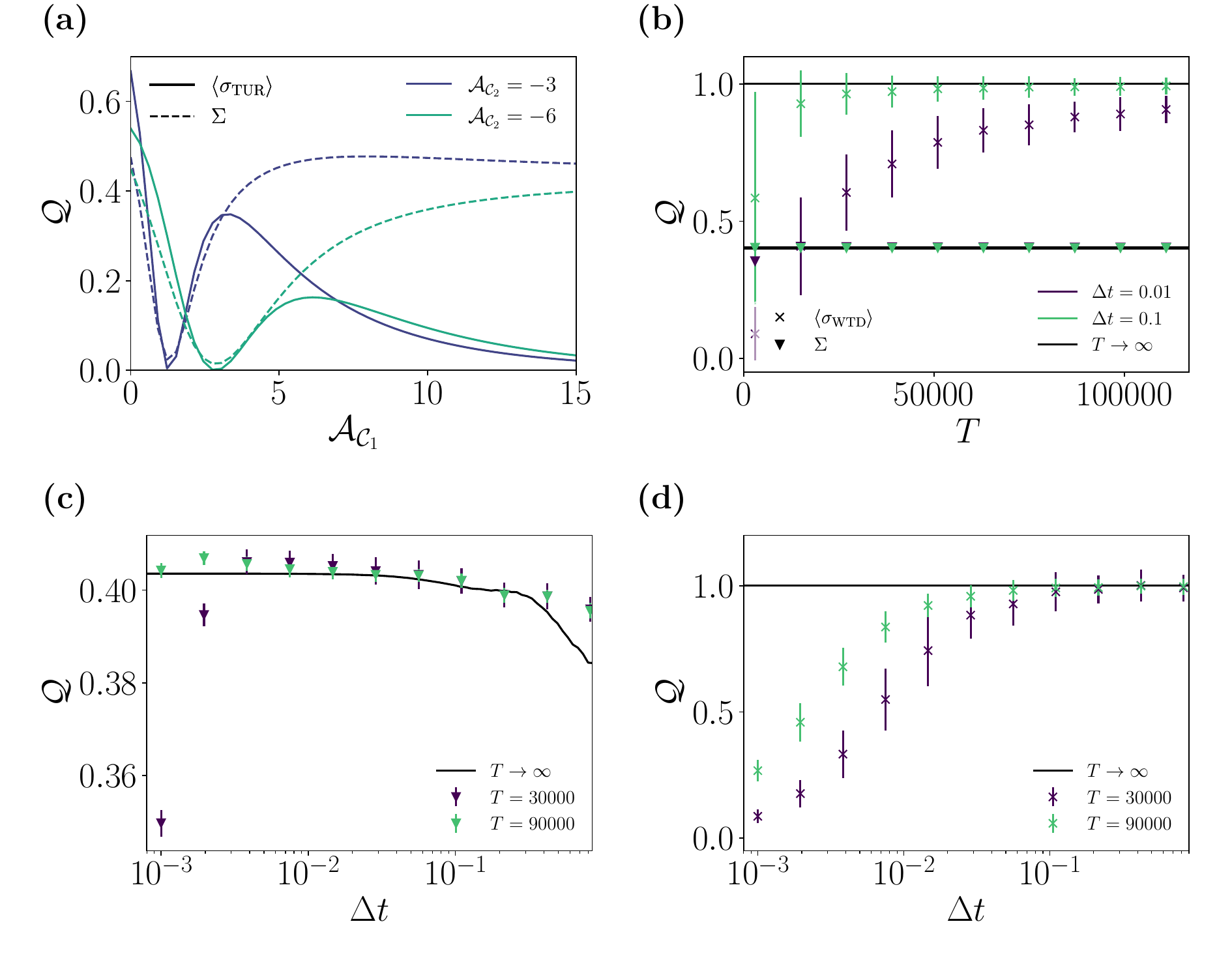}
    \caption{Effect of limited temporal and spatial resolution and finite statistics in the observation of the four-state Markov network from \autoref{Fig:SETUP_Estimation}. Only the transitions $(12),\,(21),\,(43)$ and $(34)$ are observable. An observer may not be able to distinguish between the $(12)$ and $(43)$ ($(21)$ and $(34)$) transitions, which are both in the $(+)$ ($(-)$) transition class. (a) Comparison of the estimator based on blurred transitions $\Sigma$ (dashed line) and the TUR (full line). At low driving affinity, the TUR and $\Sigma$ perform similarly. However, for higher affinities, the TUR quality factor decreases, whereas the quality factor of $\Sigma$ increases. (b) Improvement of the estimators as observation time $T$ increases. The estimator based on resolved transitions (crosses) yields a better estimate that the estimator based on blurred transitions (triangles), except for short observation times and fine time resolution. (c) Effect of finite temporal resolution and statistics on $\Sigma$. At high temporal resolution, statistical convergence is not reached, reducing the performance of the estimator. Coarser resolutions perform better, but can lead to an over-estimation of entropy. In the limit $T\rightarrow\infty$ (black line), the entropy estimator depends on the time resolution and decreases for large $\Delta t$. (d) Effect of finite temporal resolution and statistics on $\braket{\sigma_{\text{WTD}}}$. As the time resolution increases, the estimator gets worse, but self-averaging reduces variance indicated by error bars. At low temporal resolution, the variance also decreases due to the lower number of bins and faster statistical convergence. When transitions are only counted with their waiting-times discarded, the full entropy production is still recovered because the remaining underlying network is topologically trivial. The specific parameters are $\kappa_1=10,\,\kappa_2=1$ in (a)-(d) and $A_{\mathcal{C}_1}=7.5, \, A_{\mathcal{C}_2}=-4.5$ in (b)-(d) for the parametrization given in \cref{SEC:Markov}. For (b)-(d), we simulate $N=150$ trajectories.}
    \label{fig:diamond}
\end{figure*}

\subsection{Simulation results}

We start by considering the observation of the four-state Markov network from \cref{Fig:SETUP_Estimation} with observable transitions $(12), (21), (34)$ and $(43)$. The network consists of two fundamental cycles $\mathcal{C}_1$ and $\mathcal{C}_1$ with three edges, driven by affinities $\mathcal{A}_{\mathcal{C}_1}$ and $\mathcal{A}_{\mathcal{C}_2}$, respectively. The forward (backward) rates in cycle $l$ are parametrized as $k^{\pm}_{l} = \kappa_l \exp{\left(\pm A_{\mathcal{C}_l}/6\right)}$. In each cycle, one of the edges can be observed, although only forward ($+$) and backward ($-$) transitions might be distinguishable. To quantify the performance of the different entropy estimators, we introduce the quality factor
\begin{equation}
    \mathcal{Q}\equiv \frac{\langle\hat{\sigma}\rangle}{\braket{\sigma}}\,,
    \label{EQ:Q}
\end{equation}
where $\langle\hat{\sigma}\rangle\in\left\{ \braket{\sigma_{\mathrm{WTD}}},\braket{\sigma_{\mathrm{TUR}}},\Sigma\right\}$. For $\mathcal{Q} = 1$, the considered bound for $\braket{\sigma}$ saturates and the corresponding entropy estimator $\braket{\hat{\sigma}}$ infers the full entropy production. If $\mathcal{Q} > 1$, $\braket{\hat{\sigma}}$ is not a thermodynamically consistent entropy estimator for the respective observation scenario. 

If the observable transitions are all distinguishable, $\braket{\sigma_{\mathrm{WTD}}}$ recovers the full entropy production regardless of affinities $A_{\mathcal{C}_1}$ and $A_{\mathcal{C}_2}$, because each subcycle contributes a visible edge \cite{PRX}. If spatial resolution is limited such that forward and backward transitions of the cycles are blurred, entropy production can be estimated by using the TUR from \cref{EQ:SETUP_TUR2} or the bound from \cref{EQ:SETUP_FiniteT7} for blurred transitions, see \cref{fig:diamond} (a). In this case, the generalized current for the TUR is the sum of the currents across each visible link. At low affinities, the estimator for blurred transitions, $\Sigma$, performs similar to the TUR-estimator $\braket{\sigma_{\mathrm{TUR}}}$. Note that, because the system is not in equilibrium ($\mathcal{A}_{\mathcal{C}_2}\neq 0$), the quality factor for the TUR is smaller than $1$ even for $\mathcal{A}_{\mathcal{C}_1}=0$. At high affinities, the quality factor for $\Sigma$ improves, whereas the quality factor for $\braket{\sigma_{\mathrm{TUR}}}$ decreases. Thus, different estimators provide a tighter lower bound for different regimes of affinity. Therefore, if the underlying affinity is unknown, different estimators should be tested to see which one provides the tightest lower bound.

\begin{figure}[bt]
    \includegraphics[width=0.5\textwidth]{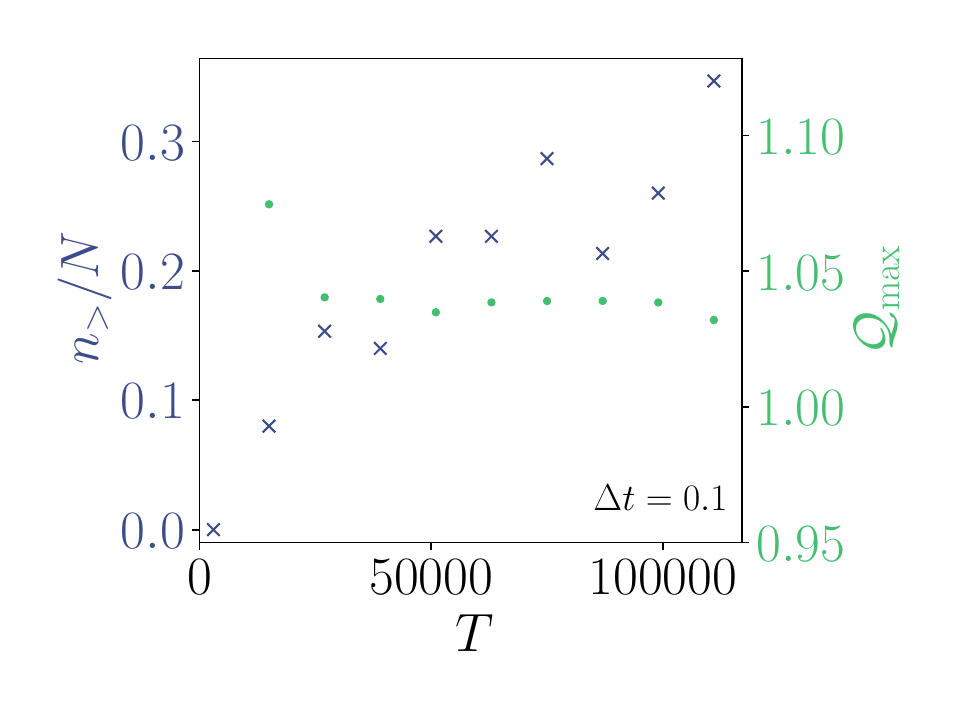}
    \caption{Overestimation of entropy production due to finite statistics. The relative occurrence of overestimation $n_>/N$ (blue crosses) increases as the mean of the estimate $\langle\sigma_{\mathrm{WTD}}\rangle$ tends towards the true value of $\langle\sigma\rangle$ with increasing observation time $T$. At the same time, the maximal overestimation $\mathcal{Q}_{\mathrm{max}}$ (green dots) decreases as the variance of the estimator (light green error bars in \cref{fig:diamond} (b)) decreases.}
    \label{fig:excursions}
\end{figure}

To investigate the influence of finite time resolution and finite statistics, we perform Gillespie simulations \cite{gillespie1977}. The waiting-time distributions are measured as empirical histograms as in \autoref{EQ:SETUP_FiniteT1} with bin width $\Delta t$. To characterize the fluctuations of the waiting-time distributions and the resulting estimators, the simulations are repeated $N$ times. One example of a fluctuating waiting-time distribution is given in \cref{Fig:SETUP_TempCG} (b). For a finite observation time, it is possible that certain events $(\tilde{kl})\overset{t_i}{\to} (\tilde{ij})$ are not observed. If this is the case, contributions for which $\psi_{(\tilde{kl})\to (\tilde{ij})}^{\Delta t}(t_i)=0$ are discarded. The resulting entropy estimates from $\left\langle\Sigma^{\Delta t,\,T}\right\rangle_N$ and $\left\langle\sigma_{\text{WTD}}^{\Delta t,\,T}\right\rangle_N$ are shown in \cref{fig:diamond} (b)-(d). In \cref{fig:diamond} (b) we investigate how the quality of these estimators scales with observation time $T$ at two different temporal resolutions. While $\left\langle\sigma_{\text{WTD}}^{\Delta t,\,T}\right\rangle_N$ achieves a better estimate, it converges more slowly towards the $T\to\infty$ limit and has a much higher variance than $\Sigma^{\Delta t,\,T}$, due to the higher number of waiting-time distributions that need to be sampled. At low observation times and high temporal resolution, this effect even leads to a better performance of the estimator based on blurred transitions. Crucially, the large variance can also lead to an overestimation of entropy production in a single run. We illustrate this further in \cref{fig:excursions} with data from \cref{fig:diamond} (b). The ratio of runs that overestimate the entropy production, $n_>/N$, increases as the mean of the estimator $\langle \sigma_{\mathrm{WTD}}\rangle$ approaches the true entropy production. Nevertheless, the maximum overestimation $\mathcal{Q}_{\mathrm{max}}$ decreases as the variance decreases.

The effect of finite temporal resolution $\Delta t$ on $\left\langle\Sigma^{\Delta t,\,T}\right\rangle_N$ and $\left\langle\sigma_{\text{WTD}}^{\Delta t,\,T}\right\rangle_N$ for a fixed value of $T$ is shown in \cref{fig:diamond} (c) and (d) respectively. In both cases, for a high time resolution, i.e., small values of $\Delta t$, entropy production is greatly underestimated. This numerical observation is caused by insufficient observation statistics for a high resolution, i.e., few events need to be sorted into a large number of bins. Thus, most of the contributions need to be discarded, since either $\psi_{(ij)\to (kl)}^{\Delta t,\,T}(t)$ or $\psi_{(\tilde{kl})\to (\tilde{ij})}^{\Delta t,\,T}(t)$ is $0$. As a consequence, increasing the length of the observation greatly benefits the estimators in this regime, as more events appear at least once. At coarser resolution, the histograms converge towards the true distribution more quickly. Here, fluctuations can also lead to an overestimation of the entropy bound. In this specific setup of the four-state Markov network, for $T\to\infty$, a finite time resolution $\Delta t$ does not affect the estimator based on resolved transitions but results in a decrease in the estimator based on blurred transitions for large $\Delta t$, see the black lines in \cref{fig:diamond} (c) and (d). For resolved transitions, this observation is not surprising. If one edge of each fundamental cycle can be observed, the estimator recovers the full entropy production independent of the time resolution of the observation \cite{PRX,harunari2022}. For blurred transitions, this observation implies that there is a trade-off between incomplete sampling at high resolutions and the lower quality factor at low resolutions. The behavior of the estimator in \cref{fig:diamond} (c) is reminiscent of \cite{luce23}, where it was found in a different setup, with a similar temporal coarse-graining. \cref{fig:diamond} (d) additionally illustrates another property of waiting-time based estimators. Each summand, i.e., each bin of each waiting-time distribution, in \cref{EQ:SETUP_FiniteT3} is a fluctuating quantity that depends on the specific realization of the observed trajectory. Since \cref{EQ:SETUP_FiniteT3} is a sum over these fluctuating quantities, a self-averaging effect takes place. This effect becomes more pronounced, as the number of summands, i.e., the number of bins, increases. Thus, even though the estimation of the mean is worse for a higher time resolution, its variance is lower. At the same time, a lower temporal resolution leads to faster statistical convergence in which the relative fluctuations decrease, such that the variance also decreases in this limit. Thus, variance is highest at intermediate temporal resolution.
\subsection{Role of different contributions to $\Sigma$}

\begin{figure}
    \includegraphics[width=0.5\textwidth]{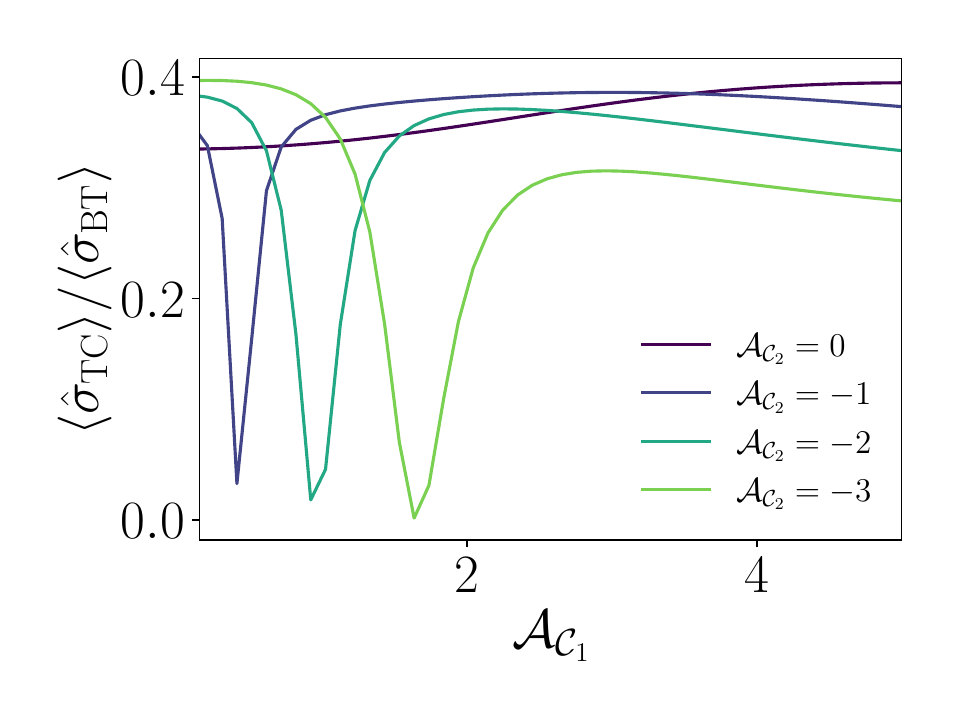}
    \caption{Contributions of $\braket{\hat{\sigma}_{\mathrm{BT}}}$ and $\braket{\hat{\sigma}_{\mathrm{TC}}}$ to $\Sigma$ as a function of $\mathcal{A}_{\mathcal{C}_1}$ for the four-state network. For $\mathcal{A}_{\mathcal{C}_1}\neq 0$, there is a sharp drop in the ratio $\braket{\hat{\sigma}_{\mathrm{TC}}}/\braket{\hat{\sigma}_{\mathrm{BT}}}$, which corresponds to the dip in the quality factor in \cref{fig:diamond} (b).}
    \label{fig:TC_vs_BT}
\end{figure}

The estimator based on blurred transitions $\Sigma$ includes the two contributions $\braket{\hat{\sigma}_{\mathrm{BT}}}$ and $\braket{\hat{\sigma}_{\mathrm{TC}}}$. To investigate whether there is an order relation between these two quantities, we analyze their ratio $\braket{\hat{\sigma}_{\mathrm{TC}}}/\braket{\hat{\sigma}_{\mathrm{BT}}}$ for different cycle affinities. As shown in \autoref{fig:TC_vs_BT} this ratio is always smaller than one. Furthermore, there is a prominent dip for $\mathcal{A}_{\mathcal{C}_2}\neq 0$ which corresponds to the dip in \cref{fig:diamond} (b). This dip is absent for $\mathcal{A}_{\mathcal{C}_2} = 0$, which suggests that the dip occurs when the affinities of the subcycles cancel along the diagonal link, i.e., when $\mathcal{A}_{\mathcal{C}_1}\simeq -\mathcal{A}_{\mathcal{C}_2}$.

Since $\braket{\hat{\sigma}_{\mathrm{TC}}}/\braket{\hat{\sigma}_{\mathrm{BT}}} < 1$ holds for the range of affinities considered here, $\braket{\hat{\sigma}_{\mathrm{BT}}}$ is the dominant contribution to $\Sigma$ for the corresponding rates in the four-state network. This result is in contrast to the results in the recent Ref. \cite{harunari2024_2}, where it was assumed and observed that the dominant contribution to $\Sigma$ is $\braket{\hat{\sigma}_{\mathrm{TC}}}$. Therefore, we conclude that there is no order relation between $\braket{\hat{\sigma}_{\mathrm{TC}}}$ and $\braket{\hat{\sigma}_{\mathrm{BT}}}$, in general.

\section{Michaelis-Menten reaction scheme}
\label{SEC:Micha}
\subsection{Model}

\begin{figure}[bt]
    \includegraphics[width=0.6\linewidth]{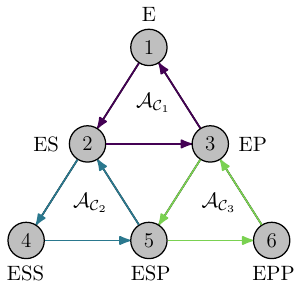}
    \caption[Setup figure 2]{Six-state Markov network for reversible two-substrate Michaelis-Menten kinetics with intermediate state. The possible reaction paths correspond to the different cycles $\mathcal{C}_1$ (dark color), $\mathcal{C}_2$ (intermediate color) and $\mathcal{C}_3$ (bright color) with affinities $\mathcal{A}_{\mathcal{C}_1} = \mathcal{A}_{\mathcal{C}_2} = \mathcal{A}_{\mathcal{C}_3} = \mathcal{A}$.}
\label{Fig:SETUP_Michaelis}
\end{figure}

One paradigm of theoretical enzyme kinetics is the Michaelis-Menten reaction mechanism \cite{fersht2002,ge2012,lervik2015,wachtel2018}. Assuming reversible creation of products, the basic realization of this mechanism is given by
\begin{equation}
    \ce{E + S}\ce{<=>[{k_+^{\mathrm{S}}}][{k_{-}^{\mathrm{S}}}] ES}\ce{<=>[{k_-^{\mathrm{P}}}][{k_{+}^{\mathrm{P}}}] E + P}
    \label{EQ:SETUP_MM1}
\end{equation}
where E is the enzyme, S is a substrate molecule and P is a product molecule and $k_{\pm}^{\text{S/P}}$ are the transition rates of the reactions. We assume that substrate and product molecules are provided at constant concentrations by chemical reservoirs. Thus, the transition rates $k_+^{\mathrm{S}}$ and $k_+^{\mathrm{P}}$ are directly proportional to the respective concentrations. The state of the enzyme, and therefore the reaction scheme in \autoref{EQ:SETUP_MM1} can be described by a Markov network with two states and four different transitions. Based on \autoref{EQ:SETUP_Aff}, the affinity of this unicyclic Markov network is given by
\begin{equation}
    \mathcal{A} = \ln\frac{k_+^{\mathrm{S}}k_-^{\mathrm{P}}}{k_-^{\mathrm{S}}k_+^{\mathrm{P}}}.
    \label{EQ:SETUP_MM_Aff1}
\end{equation}
From a chemical perspective, the system can be driven out of equilibrium by a difference in chemical potential $\Delta\mu$ between the reservoirs which implies
\begin{equation}
    \mathcal{A} = \Delta\mu = \mu_{\text{S}} - \mu_{\text{P}} \neq 0
    \label{EQ:SETUP_MM_Aff2}
\end{equation}
for the affinity of the corresponding cycle.

Often, enzymatic networks are more complex than the unicyclic network defined by \cref{EQ:SETUP_MM1}. A more sophisticated model was discussed in Ref. \cite{barato2015_2}, in which an enzyme E can bind two substrate molecules with an additional intermediate state EP in the reaction mechanism. Depending on the number of bounded substrate molecules S, this enzyme can undergo the reactions
\begin{equation}
    \ce{E + S}\ce{<=>[{k_{1+}^{\mathrm{S}}}][{k_{1-}^{\mathrm{S}}}] ES}\ce{<=>[{k_{1+}^{\mathrm{I}}}][{k_{1-}^{\mathrm{I}}}] EP}\ce{<=>[{k_{1-}^{\mathrm{P}}}][{k_{1+}^{\mathrm{P}}}] E + P},
    \label{EQ:SETUP_MM2}
\end{equation}
\begin{equation}
    \ce{ES + S}\ce{<=>[{k_{2+}^{\mathrm{S}}}][{k_{2-}^{\mathrm{S}}}] ESS}\ce{<=>[{k_{2+}^{\mathrm{I}}}][{k_{2-}^{\mathrm{I}}}] ESP}\ce{<=>[{k_{2-}^{\mathrm{P}}}][{k_{2+}^{\mathrm{P}}}] ES + P}
    \label{EQ:SETUP_MM3}
\end{equation}
and
\begin{equation}
    \ce{EP + S}\ce{<=>[{k_{3+}^{\mathrm{S}}}][{k_{3-}^{\mathrm{S}}}] EPS}\ce{<=>[{k_{3+}^{\mathrm{I}}}][{k_{3-}^{\mathrm{I}}}] EPP}\ce{<=>[{k_{3-}^{\mathrm{P}}}][{k_{3+}^{\mathrm{P}}}] EP + P}.
    \label{EQ:SETUP_MM4}
\end{equation}

\begin{figure*}[bt]
    \includegraphics[width=0.95\textwidth]{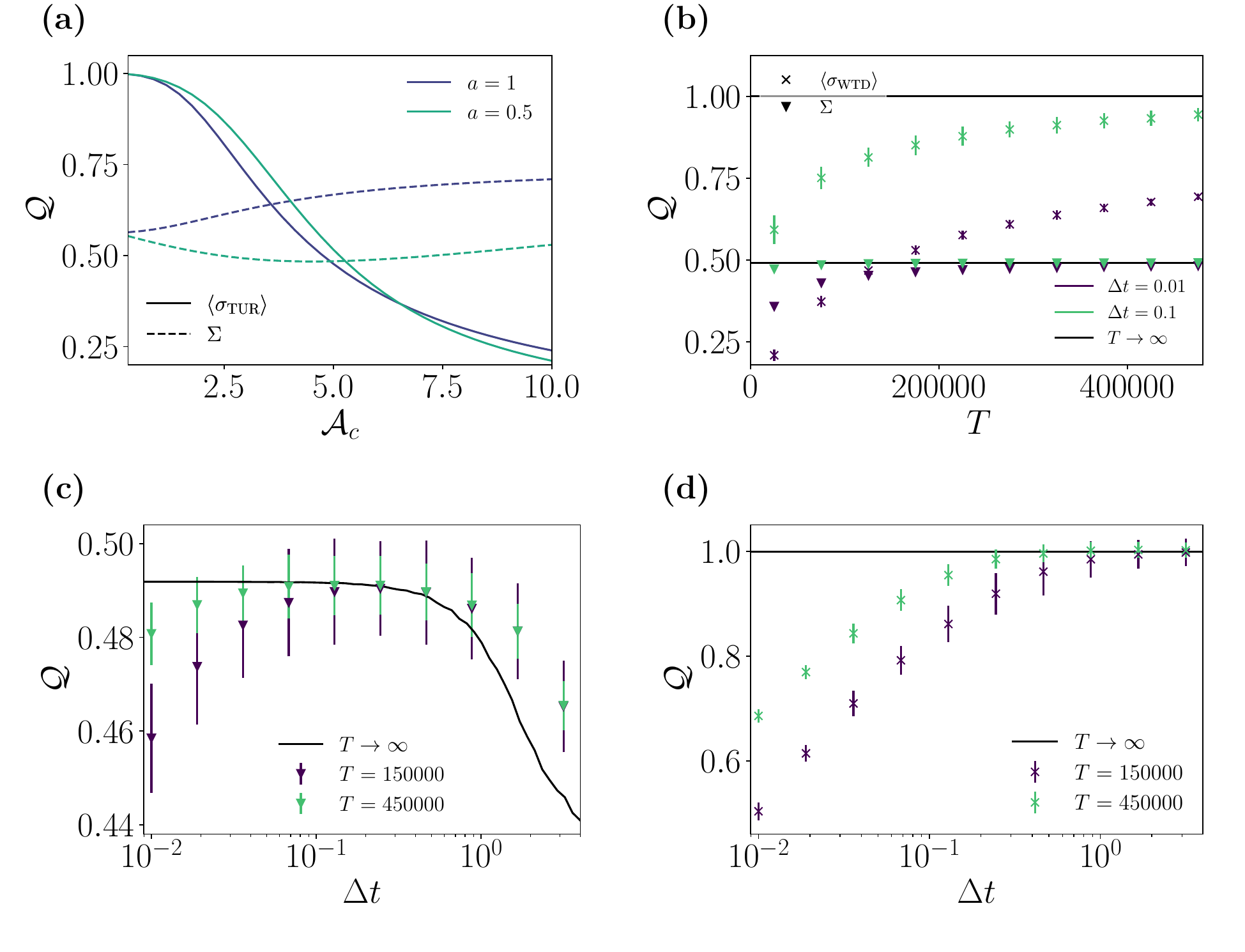}
    \caption{The effect of limited temporal and spatial resolution and finite statistics for the observation of substrate binding and release in the Michaelis-Menten reaction scheme from \autoref{Fig:SETUP_Michaelis}. (a) Comparison of $\Sigma$ and $\braket{\sigma_{\mathrm{TUR}}}$. At low driving affinity, the TUR (full lines) significantly outperforms $\Sigma$ (dashed lines), but gets worse at higher affinities while $\Sigma$ increases. (b) Dependence of the estimators on observation time $T$. At shorter observation times, the estimator based on blurred transitions (triangles) outperforms the estimator based on resolved transitions (crosses) at lower spatial resolution due to faster statistical convergence towards its limit for $T\to\infty$ (black lines). (c) Effect of finite temporal resolution on $\Sigma$. At low time resolution, i.e., for high values of $\Delta t$, less of the entropy production can be recovered in the limit $T\to\infty$, resulting in a trade-off between statistical convergence and a worse bound. (d) Effect of finite statistics on $\braket{\sigma_{\text{WTD}}}$. As the time resolution increases, the estimator gets worse. Self-averaging decreases the variance at high resolutions and statistical convergence decreases it at low resolutions. As the time resolution decreases, overestimation of entropy production occurs more often, as the error bars extend beyond $\mathcal{Q}=1$. In (b)-(d) we set $A_{\mathcal{C}_i}=3$ and $a=0.5$ and simulate $N=100$ trajectories.}
    \label{fig:michaelis}
\end{figure*}

Interpreting the six possible configurations as states of a Markov network, as shown in \cref{Fig:SETUP_Michaelis}, each reaction path corresponds to a three-state cycle $\mathcal{C}_i$ with affinity
\begin{equation}
    \mathcal{A}_{\mathcal{C}_i} = \ln\frac{k_{i+}^{\mathrm{S}}k_{i+}^{\mathrm{I}}k_{i-}^{\mathrm{P}}}{k_{i-}^{\mathrm{S}}k_{i-}^{\mathrm{I}}k_{i+}^{\mathrm{P}}}.
    \label{EQ:SETUP_MM_Aff3}
\end{equation}
Since each cycle is fueled by the same reservoirs, all affinities from \autoref{EQ:SETUP_MM_Aff3} are identical and correspond to the difference in chemical potentials $\Delta\mu$ between these reservoirs, i.e.
\begin{equation}
    \mathcal{A}_{\mathcal{C}_1} = \mathcal{A}_{\mathcal{C}_2} = \mathcal{A}_{\mathcal{C}_3} = \mathcal{A} = \Delta\mu = \mu_{\text{S}} - \mu_{\text{P}}.
    \label{EQ:SETUP_MM_Aff4}
\end{equation}
Note that, by construction, the additional cycle $\mathcal{C}_4=2532$ has affinity zero because it includes only binding and unbinding of molecules and no chemical conversion.

The rates in the Markov network are parametrized by $k_{i\pm}^{\mathrm{S}} = \exp\left(\pm a \mathcal{A}_{\mathcal{C}_i}/2\right)$, $k_{i\pm}^{\mathrm{P}}= \exp\left(\pm (1-a) \mathcal{A}_{\mathcal{C}_i}/2\right)$ and $k_{i\pm}^{\mathrm{I}}=1$. The parameter $a$ measures how the affinity in each cycle is split between the substrate binding and product releasing process. Experimentally, this parameter can be controlled by varying the substrate and product concentration.

\subsection{Simulation results}
As observation scenario for the described augmented Michaelis-Menten reaction scheme, we assume that an external observer can only register the binding and unbinding of substrate molecules, which is equivalent to observing the transitions $(12),(21),(24),(42),(35)$ and $(53)$ in the corresponding Markov network. If an observer cannot tell which other proteins are bound to the enzyme, i.e., in which of the cycles the reaction takes place, his observation has finite spatial resolution and thus blurred transitions are registered. In \cref{fig:michaelis}, we analyze the quality factors of the different estimators for the described observation, which are defined analogously to \cref{EQ:Q}. For the observation of blurred transitions, \cref{fig:michaelis} (a) shows the quality factors of $\Sigma$ and $\langle\sigma_{\mathrm{TUR}}\rangle$. For the TUR, the generalized current is the sum of the currents across the three visible links. If the six transitions can be resolved individually, $\sigma_{\mathrm{WTD}}$ again recovers the full entropy production due to the topology of the network and observed links, for which one link of each cycle is observed. Close to equilibrium, the TUR saturates and significantly beats $\Sigma$, but loses out at high affinities far from equilibrium. For finite statistic and finite time resolution, the entropy estimates from the waiting-time based estimators are shown in \cref{fig:michaelis} (b)-(d). In \cref{fig:michaelis} (b) we investigate how the increase in observation time affects both estimators. Again, we find that the blurred estimator performs significantly better at low observation times, where the lower spatial resolution is an advantage due to faster convergence. In \cref{fig:michaelis} (c) and (d), we show the effect of finite temporal resolution $\Delta t$ at fixed values of $T$ on $\Sigma$ and $\langle\sigma_{\mathrm{WTD}}\rangle$, respectively. As in \cref{fig:diamond} (c) and (d), there is a trade-off between better statistical convergence and a worse bound in the limit $T\to\infty$ at low resolutions for $\Sigma$. Furthermore, there is a trade-off in the variance between the self-averaging effect at high resolution and statistical convergence at low resolution for $\braket{\sigma_{\mathrm{WTD}}}$. 
\section{Conclusion}
\label{SEC:Conclusion}

In this work, we have studied the general effects that arise when entropy estimators are applied to observations with limited spatial and temporal resolution and finite statistics. Furthermore, we have shown that the empirical waiting-time distributions have Gaussian fluctuations in general. To illustrate these imperfect observation scenarios, we chose two paradigmatic examples: a four-state network and an augmented realization of the Michaelis-Menten reaction mechanism. For scenarios with limited spatial resolution, an estimator based on the TUR is better suited for low cycle affinities, i.e. close to equilibrium, whereas the waiting-time based estimator for blurred transitions performs better for high cycle affinities. For a finite amount of observation data, a lower time resolution can be beneficial for the quality of the considered estimators. However, at lower time resolution, even a statistically converged estimator may recover less entropy production than at higher time resolution. In general, in scenarios with finite statistics, fluctuations may lead to an overestimation of entropy production and empirical quantities fail to provide a strict lower bound on entropy production. For waiting-time based estimators, based on the observation of either resolved or blurred transitions, a large number of bins for the empirical waiting-time distributions introduces a self-averaging effect that reduces the statistical variance of these estimators. In addition, the estimator based on blurred transitions achieves faster convergence due to the lower number of waiting-time distributions it is based on.

Although these effects emerge in two specific examples, their fundamental nature hints at broader implications for inferring entropy production based on imperfect observations. In particular, the observed overestimation of the entropy production for finite observation statistics calls for the quantification of the overestimation and a formulation of consistent entropy estimators for imperfect observation scenarios. Therefore, a conceptual generalization aiming at general entropy estimators for imperfect observations based on our operationally motivated results might be a promising next step. In particular, the temporal coarse-graining we have considered still assumes that all visible transitions are registered. It would be promising to study other kinds of temporal coarse-graining, in which, for example, only a single event can be registered in each time interval. Similarly, the effects of finite statistics on other estimators designed for different types of spatial coarse-graining, such as state-lumping, could be studied. Nevertheless, the observation scenarios considered here are a first step in describing imperfect observations in experiments. Thus, it might be possible to observe the discussed effects, for example the self-averaging effect, also for experimental data. Furthermore, we expect to observe similar results in imperfect observations of systems with continuous dynamics and beyond the steady state, where statistical convergence is even harder to achieve. 

\begin{acknowledgments}
We thank Julius Degünther and Jann van der Meer for fruitful discussions.
\end{acknowledgments}



%


\begin{thebibliography}{101}%
\makeatletter
\providecommand \@ifxundefined [1]{%
 \@ifx{#1\undefined}
}%
\providecommand \@ifnum [1]{%
 \ifnum #1\expandafter \@firstoftwo
 \else \expandafter \@secondoftwo
 \fi
}%
\providecommand \@ifx [1]{%
 \ifx #1\expandafter \@firstoftwo
 \else \expandafter \@secondoftwo
 \fi
}%
\providecommand \natexlab [1]{#1}%
\providecommand \enquote  [1]{``#1''}%
\providecommand \bibnamefont  [1]{#1}%
\providecommand \bibfnamefont [1]{#1}%
\providecommand \citenamefont [1]{#1}%
\providecommand \href@noop [0]{\@secondoftwo}%
\providecommand \href [0]{\begingroup \@sanitize@url \@href}%
\providecommand \@href[1]{\@@startlink{#1}\@@href}%
\providecommand \@@href[1]{\endgroup#1\@@endlink}%
\providecommand \@sanitize@url [0]{\catcode `\\12\catcode `\$12\catcode `\&12\catcode `\#12\catcode `\^12\catcode `\_12\catcode `\%12\relax}%
\providecommand \@@startlink[1]{}%
\providecommand \@@endlink[0]{}%
\providecommand \url  [0]{\begingroup\@sanitize@url \@url }%
\providecommand \@url [1]{\endgroup\@href {#1}{\urlprefix }}%
\providecommand \urlprefix  [0]{URL }%
\providecommand \Eprint [0]{\href }%
\providecommand \doibase [0]{https://doi.org/}%
\providecommand \selectlanguage [0]{\@gobble}%
\providecommand \bibinfo  [0]{\@secondoftwo}%
\providecommand \bibfield  [0]{\@secondoftwo}%
\providecommand \translation [1]{[#1]}%
\providecommand \BibitemOpen [0]{}%
\providecommand \bibitemStop [0]{}%
\providecommand \bibitemNoStop [0]{.\EOS\space}%
\providecommand \EOS [0]{\spacefactor3000\relax}%
\providecommand \BibitemShut  [1]{\csname bibitem#1\endcsname}%
\let\auto@bib@innerbib\@empty
\bibitem [{\citenamefont {Sekimoto}(2010)}]{Sekimoto2010}%
  \BibitemOpen
  \bibfield  {author} {\bibinfo {author} {\bibfnamefont {K.}~\bibnamefont {Sekimoto}},\ }\href {https://doi.org/10.1007/978-3-642-05411-2} {\emph {\bibinfo {title} {Stochastic Energetics}}},\ Lecture {Notes} in {Physics}\ (\bibinfo  {publisher} {Springer Berlin Heidelberg},\ \bibinfo {address} {Berlin, Heidelberg},\ \bibinfo {year} {2010})\BibitemShut {NoStop}%
\bibitem [{\citenamefont {Jarzynski}(2011)}]{jarzynski2011}%
  \BibitemOpen
  \bibfield  {author} {\bibinfo {author} {\bibfnamefont {C.}~\bibnamefont {Jarzynski}},\ }\href {https://doi.org/10.1146/annurev-conmatphys-062910-140506} {\bibfield  {journal} {\bibinfo  {journal} {Annu. Rev. Condens. Matter Phys.}\ }\textbf {\bibinfo {volume} {2}},\ \bibinfo {pages} {329} (\bibinfo {year} {2011})}\BibitemShut {NoStop}%
\bibitem [{\citenamefont {Seifert}(2012)}]{seifert2012}%
  \BibitemOpen
  \bibfield  {author} {\bibinfo {author} {\bibfnamefont {U.}~\bibnamefont {Seifert}},\ }\href {https://doi.org/10.1088/0034-4885/75/12/126001} {\bibfield  {journal} {\bibinfo  {journal} {Rep. Prog. Phys.}\ }\textbf {\bibinfo {volume} {75}},\ \bibinfo {pages} {126001} (\bibinfo {year} {2012})}\BibitemShut {NoStop}%
\bibitem [{\citenamefont {Van~den Broeck}\ and\ \citenamefont {Esposito}(2015)}]{broeck2015}%
  \BibitemOpen
  \bibfield  {author} {\bibinfo {author} {\bibfnamefont {C.}~\bibnamefont {Van~den Broeck}}\ and\ \bibinfo {author} {\bibfnamefont {M.}~\bibnamefont {Esposito}},\ }\href {https://doi.org/10.1016/j.physa.2014.04.035} {\bibfield  {journal} {\bibinfo  {journal} {Phys. A: Stat. Mech. Appl.}\ }\textbf {\bibinfo {volume} {418}},\ \bibinfo {pages} {6} (\bibinfo {year} {2015})}\BibitemShut {NoStop}%
\bibitem [{\citenamefont {Schmiedl}\ and\ \citenamefont {Seifert}(2007)}]{schmiedl2007}%
  \BibitemOpen
  \bibfield  {author} {\bibinfo {author} {\bibfnamefont {T.}~\bibnamefont {Schmiedl}}\ and\ \bibinfo {author} {\bibfnamefont {U.}~\bibnamefont {Seifert}},\ }\href {https://doi.org/10.1063/1.2428297} {\bibfield  {journal} {\bibinfo  {journal} {J. Chem. Phys.}\ }\textbf {\bibinfo {volume} {126}},\ \bibinfo {pages} {044101} (\bibinfo {year} {2007})}\BibitemShut {NoStop}%
\bibitem [{\citenamefont {Ge}\ \emph {et~al.}(2012)\citenamefont {Ge}, \citenamefont {Qian},\ and\ \citenamefont {Qian}}]{ge2012}%
  \BibitemOpen
  \bibfield  {author} {\bibinfo {author} {\bibfnamefont {H.}~\bibnamefont {Ge}}, \bibinfo {author} {\bibfnamefont {M.}~\bibnamefont {Qian}},\ and\ \bibinfo {author} {\bibfnamefont {H.}~\bibnamefont {Qian}},\ }\href {https://doi.org/https://doi.org/10.1016/j.physrep.2011.09.001} {\bibfield  {journal} {\bibinfo  {journal} {Phys. Rep.}\ }\textbf {\bibinfo {volume} {510}},\ \bibinfo {pages} {87} (\bibinfo {year} {2012})}\BibitemShut {NoStop}%
\bibitem [{\citenamefont {Rao}\ and\ \citenamefont {Esposito}(2016)}]{rao2016}%
  \BibitemOpen
  \bibfield  {author} {\bibinfo {author} {\bibfnamefont {R.}~\bibnamefont {Rao}}\ and\ \bibinfo {author} {\bibfnamefont {M.}~\bibnamefont {Esposito}},\ }\href {https://doi.org/10.1103/PhysRevX.6.041064} {\bibfield  {journal} {\bibinfo  {journal} {Phys. Rev. X}\ }\textbf {\bibinfo {volume} {6}},\ \bibinfo {pages} {041064} (\bibinfo {year} {2016})}\BibitemShut {NoStop}%
\bibitem [{\citenamefont {Qian}(2000)}]{qian2000}%
  \BibitemOpen
  \bibfield  {author} {\bibinfo {author} {\bibfnamefont {H.}~\bibnamefont {Qian}},\ }\href {https://doi.org/10.1016/s0301-4622(99)00121-0} {\bibfield  {journal} {\bibinfo  {journal} {Biophys. Chem.}\ }\textbf {\bibinfo {volume} {83}},\ \bibinfo {pages} {35} (\bibinfo {year} {2000})}\BibitemShut {NoStop}%
\bibitem [{\citenamefont {Andrieux}\ and\ \citenamefont {Gaspard}(2006)}]{andrieux2006}%
  \BibitemOpen
  \bibfield  {author} {\bibinfo {author} {\bibfnamefont {D.}~\bibnamefont {Andrieux}}\ and\ \bibinfo {author} {\bibfnamefont {P.}~\bibnamefont {Gaspard}},\ }\href {https://doi.org/10.1103/PhysRevE.74.011906} {\bibfield  {journal} {\bibinfo  {journal} {Phys. Rev. E}\ }\textbf {\bibinfo {volume} {74}},\ \bibinfo {pages} {011906} (\bibinfo {year} {2006})}\BibitemShut {NoStop}%
\bibitem [{\citenamefont {Seifert}(2011)}]{seifert2011}%
  \BibitemOpen
  \bibfield  {author} {\bibinfo {author} {\bibfnamefont {U.}~\bibnamefont {Seifert}},\ }\href {https://doi.org/10.1140/epje/i2011-11026-7} {\bibfield  {journal} {\bibinfo  {journal} {Phys. J. E}\ }\textbf {\bibinfo {volume} {34}} (\bibinfo {year} {2011})}\BibitemShut {NoStop}%
\bibitem [{\citenamefont {Chowdhury}(2013)}]{chowdhury2013}%
  \BibitemOpen
  \bibfield  {author} {\bibinfo {author} {\bibfnamefont {D.}~\bibnamefont {Chowdhury}},\ }\href {https://doi.org/https://doi.org/10.1016/j.physrep.2013.03.005} {\bibfield  {journal} {\bibinfo  {journal} {Phys. Rep.}\ }\textbf {\bibinfo {volume} {529}},\ \bibinfo {pages} {1} (\bibinfo {year} {2013})}\BibitemShut {NoStop}%
\bibitem [{\citenamefont {Kolomeisky}(2015)}]{kolomeisky_book}%
  \BibitemOpen
  \bibfield  {author} {\bibinfo {author} {\bibfnamefont {A.~B.}\ \bibnamefont {Kolomeisky}},\ }\href@noop {} {\emph {\bibinfo {title} {Motor Proteins and Molecular Motors}}}\ (\bibinfo  {publisher} {CRC Press.},\ \bibinfo {address} {Boca Raton, USA},\ \bibinfo {year} {2015})\BibitemShut {NoStop}%
\bibitem [{\citenamefont {Speck}(2021)}]{speck2021}%
  \BibitemOpen
  \bibfield  {author} {\bibinfo {author} {\bibfnamefont {T.}~\bibnamefont {Speck}},\ }\href {https://doi.org/10.1063/5.0070922} {\bibfield  {journal} {\bibinfo  {journal} {J. Chem. Phys.}\ }\textbf {\bibinfo {volume} {155}},\ \bibinfo {pages} {230901} (\bibinfo {year} {2021})}\BibitemShut {NoStop}%
\bibitem [{\citenamefont {Ritort}(2006)}]{ritort2006}%
  \BibitemOpen
  \bibfield  {author} {\bibinfo {author} {\bibfnamefont {F.}~\bibnamefont {Ritort}},\ }\href {https://doi.org/10.1088/0953-8984/18/32/r01} {\bibfield  {journal} {\bibinfo  {journal} {J. Condens. Matter Phys.}\ }\textbf {\bibinfo {volume} {18}},\ \bibinfo {pages} {R531–R583} (\bibinfo {year} {2006})}\BibitemShut {NoStop}%
\bibitem [{\citenamefont {Herbert}\ \emph {et~al.}(2008)\citenamefont {Herbert}, \citenamefont {Greenleaf},\ and\ \citenamefont {Block}}]{herbert2008}%
  \BibitemOpen
  \bibfield  {author} {\bibinfo {author} {\bibfnamefont {K.~M.}\ \bibnamefont {Herbert}}, \bibinfo {author} {\bibfnamefont {W.~J.}\ \bibnamefont {Greenleaf}},\ and\ \bibinfo {author} {\bibfnamefont {S.~M.}\ \bibnamefont {Block}},\ }\href {https://doi.org/10.1146/annurev.biochem.77.073106.100741} {\bibfield  {journal} {\bibinfo  {journal} {Annu. Rev. Biochem.}\ }\textbf {\bibinfo {volume} {77}},\ \bibinfo {pages} {149–176} (\bibinfo {year} {2008})}\BibitemShut {NoStop}%
\bibitem [{\citenamefont {Veigel}\ and\ \citenamefont {Schmidt}(2011)}]{veigel2011}%
  \BibitemOpen
  \bibfield  {author} {\bibinfo {author} {\bibfnamefont {C.}~\bibnamefont {Veigel}}\ and\ \bibinfo {author} {\bibfnamefont {C.~F.}\ \bibnamefont {Schmidt}},\ }\href {https://doi.org/10.1038/nrm3062} {\bibfield  {journal} {\bibinfo  {journal} {Nat. Rev. Mol. Cell Biol.}\ }\textbf {\bibinfo {volume} {12}},\ \bibinfo {pages} {163–176} (\bibinfo {year} {2011})}\BibitemShut {NoStop}%
\bibitem [{\citenamefont {Ariga}\ \emph {et~al.}(2018)\citenamefont {Ariga}, \citenamefont {Tomishige},\ and\ \citenamefont {Mizuno}}]{ariga2018}%
  \BibitemOpen
  \bibfield  {author} {\bibinfo {author} {\bibfnamefont {T.}~\bibnamefont {Ariga}}, \bibinfo {author} {\bibfnamefont {M.}~\bibnamefont {Tomishige}},\ and\ \bibinfo {author} {\bibfnamefont {D.}~\bibnamefont {Mizuno}},\ }\href {https://doi.org/10.1103/physrevlett.121.218101} {\bibfield  {journal} {\bibinfo  {journal} {Phys. Rev. Lett.}\ }\textbf {\bibinfo {volume} {121}},\ \bibinfo {pages} {218101} (\bibinfo {year} {2018})}\BibitemShut {NoStop}%
\bibitem [{\citenamefont {Bustamante}\ and\ \citenamefont {Yan}(2022)}]{bustamante2022}%
  \BibitemOpen
  \bibfield  {author} {\bibinfo {author} {\bibfnamefont {C.}~\bibnamefont {Bustamante}}\ and\ \bibinfo {author} {\bibfnamefont {S.}~\bibnamefont {Yan}},\ }\href {https://doi.org/10.1017/s0033583522000087} {\bibfield  {journal} {\bibinfo  {journal} {Q. Rev. Biophys.}\ }\textbf {\bibinfo {volume} {55}},\ \bibinfo {pages} {e9} (\bibinfo {year} {2022})}\BibitemShut {NoStop}%
\bibitem [{\citenamefont {Mehl}\ \emph {et~al.}(2012)\citenamefont {Mehl}, \citenamefont {Lander}, \citenamefont {Bechinger}, \citenamefont {Blickle},\ and\ \citenamefont {Seifert}}]{mehl2012}%
  \BibitemOpen
  \bibfield  {author} {\bibinfo {author} {\bibfnamefont {J.}~\bibnamefont {Mehl}}, \bibinfo {author} {\bibfnamefont {B.}~\bibnamefont {Lander}}, \bibinfo {author} {\bibfnamefont {C.}~\bibnamefont {Bechinger}}, \bibinfo {author} {\bibfnamefont {V.}~\bibnamefont {Blickle}},\ and\ \bibinfo {author} {\bibfnamefont {U.}~\bibnamefont {Seifert}},\ }\href {https://doi.org/10.1103/PhysRevLett.108.220601} {\bibfield  {journal} {\bibinfo  {journal} {Phys. Rev. Lett.}\ }\textbf {\bibinfo {volume} {108}},\ \bibinfo {pages} {220601} (\bibinfo {year} {2012})}\BibitemShut {NoStop}%
\bibitem [{\citenamefont {Bo}\ and\ \citenamefont {Celani}(2014)}]{bo2014}%
  \BibitemOpen
  \bibfield  {author} {\bibinfo {author} {\bibfnamefont {S.}~\bibnamefont {Bo}}\ and\ \bibinfo {author} {\bibfnamefont {A.}~\bibnamefont {Celani}},\ }\href {https://doi.org/10.1007/s10955-014-0922-1} {\bibfield  {journal} {\bibinfo  {journal} {J. Stat. Phys.}\ }\textbf {\bibinfo {volume} {154}},\ \bibinfo {pages} {1325} (\bibinfo {year} {2014})}\BibitemShut {NoStop}%
\bibitem [{\citenamefont {Bo}\ and\ \citenamefont {Celani}(2017)}]{bo2017}%
  \BibitemOpen
  \bibfield  {author} {\bibinfo {author} {\bibfnamefont {S.}~\bibnamefont {Bo}}\ and\ \bibinfo {author} {\bibfnamefont {A.}~\bibnamefont {Celani}},\ }\href {https://doi.org/https://doi.org/10.1016/j.physrep.2016.12.003} {\bibfield  {journal} {\bibinfo  {journal} {Phys. Rep.}\ }\textbf {\bibinfo {volume} {670}},\ \bibinfo {pages} {1} (\bibinfo {year} {2017})}\BibitemShut {NoStop}%
\bibitem [{\citenamefont {Uhl}\ \emph {et~al.}(2018)\citenamefont {Uhl}, \citenamefont {Pietzonka},\ and\ \citenamefont {Seifert}}]{uhl2018}%
  \BibitemOpen
  \bibfield  {author} {\bibinfo {author} {\bibfnamefont {M.}~\bibnamefont {Uhl}}, \bibinfo {author} {\bibfnamefont {P.}~\bibnamefont {Pietzonka}},\ and\ \bibinfo {author} {\bibfnamefont {U.}~\bibnamefont {Seifert}},\ }\href {https://doi.org/10.1088/1742-5468/aaa78b} {\bibfield  {journal} {\bibinfo  {journal} {J. Stat. Mech.}\ ,\ \bibinfo {pages} {023203}} (\bibinfo {year} {2018})}\BibitemShut {NoStop}%
\bibitem [{\citenamefont {Lucente}\ \emph {et~al.}(2022)\citenamefont {Lucente}, \citenamefont {Baldassarri}, \citenamefont {Puglisi}, \citenamefont {Vulpiani},\ and\ \citenamefont {Viale}}]{lucente2022}%
  \BibitemOpen
  \bibfield  {author} {\bibinfo {author} {\bibfnamefont {D.}~\bibnamefont {Lucente}}, \bibinfo {author} {\bibfnamefont {A.}~\bibnamefont {Baldassarri}}, \bibinfo {author} {\bibfnamefont {A.}~\bibnamefont {Puglisi}}, \bibinfo {author} {\bibfnamefont {A.}~\bibnamefont {Vulpiani}},\ and\ \bibinfo {author} {\bibfnamefont {M.}~\bibnamefont {Viale}},\ }\href {https://doi.org/10.1103/PhysRevResearch.4.043103} {\bibfield  {journal} {\bibinfo  {journal} {Phys. Rev. Res.}\ }\textbf {\bibinfo {volume} {4}},\ \bibinfo {pages} {043103} (\bibinfo {year} {2022})}\BibitemShut {NoStop}%
\bibitem [{\citenamefont {Esposito}(2012)}]{esposito2012}%
  \BibitemOpen
  \bibfield  {author} {\bibinfo {author} {\bibfnamefont {M.}~\bibnamefont {Esposito}},\ }\href {https://doi.org/10.1103/PhysRevE.85.041125} {\bibfield  {journal} {\bibinfo  {journal} {Phys. Rev. E}\ }\textbf {\bibinfo {volume} {85}},\ \bibinfo {pages} {041125} (\bibinfo {year} {2012})}\BibitemShut {NoStop}%
\bibitem [{\citenamefont {Rahav}\ and\ \citenamefont {Jarzynski}(2007)}]{rahav2007}%
  \BibitemOpen
  \bibfield  {author} {\bibinfo {author} {\bibfnamefont {S.}~\bibnamefont {Rahav}}\ and\ \bibinfo {author} {\bibfnamefont {C.}~\bibnamefont {Jarzynski}},\ }\href {https://doi.org/10.1088/1742-5468/2007/09/p09012} {\bibfield  {journal} {\bibinfo  {journal} {J. Stat. Mech.}\ ,\ \bibinfo {pages} {P09012}} (\bibinfo {year} {2007})}\BibitemShut {NoStop}%
\bibitem [{\citenamefont {Gomez-Marin}\ \emph {et~al.}(2008)\citenamefont {Gomez-Marin}, \citenamefont {Parrondo},\ and\ \citenamefont {Van~den Broeck}}]{gomez-marin2008}%
  \BibitemOpen
  \bibfield  {author} {\bibinfo {author} {\bibfnamefont {A.}~\bibnamefont {Gomez-Marin}}, \bibinfo {author} {\bibfnamefont {J.~M.~R.}\ \bibnamefont {Parrondo}},\ and\ \bibinfo {author} {\bibfnamefont {C.}~\bibnamefont {Van~den Broeck}},\ }\href {https://doi.org/10.1103/PhysRevE.78.011107} {\bibfield  {journal} {\bibinfo  {journal} {Phys. Rev. E}\ }\textbf {\bibinfo {volume} {78}},\ \bibinfo {pages} {011107} (\bibinfo {year} {2008})}\BibitemShut {NoStop}%
\bibitem [{\citenamefont {Polettini}\ and\ \citenamefont {Esposito}(2017)}]{polettini2017}%
  \BibitemOpen
  \bibfield  {author} {\bibinfo {author} {\bibfnamefont {M.}~\bibnamefont {Polettini}}\ and\ \bibinfo {author} {\bibfnamefont {M.}~\bibnamefont {Esposito}},\ }\href {https://doi.org/10.1103/PhysRevLett.119.240601} {\bibfield  {journal} {\bibinfo  {journal} {Phys. Rev. Lett.}\ }\textbf {\bibinfo {volume} {119}},\ \bibinfo {pages} {240601} (\bibinfo {year} {2017})}\BibitemShut {NoStop}%
\bibitem [{\citenamefont {Bisker}\ \emph {et~al.}(2017)\citenamefont {Bisker}, \citenamefont {Polettini}, \citenamefont {Gingrich},\ and\ \citenamefont {Horowitz}}]{bisker2017}%
  \BibitemOpen
  \bibfield  {author} {\bibinfo {author} {\bibfnamefont {G.}~\bibnamefont {Bisker}}, \bibinfo {author} {\bibfnamefont {M.}~\bibnamefont {Polettini}}, \bibinfo {author} {\bibfnamefont {T.~R.}\ \bibnamefont {Gingrich}},\ and\ \bibinfo {author} {\bibfnamefont {J.~M.}\ \bibnamefont {Horowitz}},\ }\href {https://doi.org/10.1088/1742-5468/aa8c0d} {\bibfield  {journal} {\bibinfo  {journal} {J. Stat. Mech.}\ }\textbf {\bibinfo {volume} {2017}},\ \bibinfo {pages} {093210} (\bibinfo {year} {2017})}\BibitemShut {NoStop}%
\bibitem [{\citenamefont {Pigolotti}\ and\ \citenamefont {Vulpiani}(2008)}]{pigolotti2008}%
  \BibitemOpen
  \bibfield  {author} {\bibinfo {author} {\bibfnamefont {S.}~\bibnamefont {Pigolotti}}\ and\ \bibinfo {author} {\bibfnamefont {A.}~\bibnamefont {Vulpiani}},\ }\href {https://doi.org/10.1063/1.2907242} {\bibfield  {journal} {\bibinfo  {journal} {J. Chem. Phys.}\ }\textbf {\bibinfo {volume} {128}},\ \bibinfo {pages} {154114} (\bibinfo {year} {2008})}\BibitemShut {NoStop}%
\bibitem [{\citenamefont {Puglisi}\ \emph {et~al.}(2010)\citenamefont {Puglisi}, \citenamefont {Pigolotti}, \citenamefont {Rondoni},\ and\ \citenamefont {Vulpiani}}]{puglisi2010}%
  \BibitemOpen
  \bibfield  {author} {\bibinfo {author} {\bibfnamefont {A.}~\bibnamefont {Puglisi}}, \bibinfo {author} {\bibfnamefont {S.}~\bibnamefont {Pigolotti}}, \bibinfo {author} {\bibfnamefont {L.}~\bibnamefont {Rondoni}},\ and\ \bibinfo {author} {\bibfnamefont {A.}~\bibnamefont {Vulpiani}},\ }\href {https://doi.org/10.1088/1742-5468/2010/05/p05015} {\bibfield  {journal} {\bibinfo  {journal} {J. Stat. Mech.}\ }\textbf {\bibinfo {volume} {2010}},\ \bibinfo {pages} {P05015} (\bibinfo {year} {2010})}\BibitemShut {NoStop}%
\bibitem [{\citenamefont {Teza}\ and\ \citenamefont {Stella}(2020)}]{teza2020}%
  \BibitemOpen
  \bibfield  {author} {\bibinfo {author} {\bibfnamefont {G.}~\bibnamefont {Teza}}\ and\ \bibinfo {author} {\bibfnamefont {A.~L.}\ \bibnamefont {Stella}},\ }\href {https://doi.org/10.1103/PhysRevLett.125.110601} {\bibfield  {journal} {\bibinfo  {journal} {Phys. Rev. Lett.}\ }\textbf {\bibinfo {volume} {125}},\ \bibinfo {pages} {110601} (\bibinfo {year} {2020})}\BibitemShut {NoStop}%
\bibitem [{\citenamefont {Skinner}\ and\ \citenamefont {Dunkel}(2021{\natexlab{a}})}]{skinner2021_1}%
  \BibitemOpen
  \bibfield  {author} {\bibinfo {author} {\bibfnamefont {D.~J.}\ \bibnamefont {Skinner}}\ and\ \bibinfo {author} {\bibfnamefont {J.}~\bibnamefont {Dunkel}},\ }\href {https://www.pnas.org/content/118/18/e2024300118} {\bibfield  {journal} {\bibinfo  {journal} {Proc. Natl. Acad. Sci.}\ }\textbf {\bibinfo {volume} {118}},\ \bibinfo {pages} {e2024300118} (\bibinfo {year} {2021}{\natexlab{a}})}\BibitemShut {NoStop}%
\bibitem [{\citenamefont {Ehrich}(2021)}]{ehrich2021}%
  \BibitemOpen
  \bibfield  {author} {\bibinfo {author} {\bibfnamefont {J.}~\bibnamefont {Ehrich}},\ }\href {https://doi.org/10.1088/1742-5468/ac150e} {\bibfield  {journal} {\bibinfo  {journal} {J. Stat. Mech.}\ }\textbf {\bibinfo {volume} {2021}},\ \bibinfo {pages} {083214} (\bibinfo {year} {2021})}\BibitemShut {NoStop}%
\bibitem [{\citenamefont {Nitzan}\ \emph {et~al.}(2023)\citenamefont {Nitzan}, \citenamefont {Ghosal},\ and\ \citenamefont {Bisker}}]{nitzan2023}%
  \BibitemOpen
  \bibfield  {author} {\bibinfo {author} {\bibfnamefont {E.}~\bibnamefont {Nitzan}}, \bibinfo {author} {\bibfnamefont {A.}~\bibnamefont {Ghosal}},\ and\ \bibinfo {author} {\bibfnamefont {G.}~\bibnamefont {Bisker}},\ }\href {https://doi.org/10.1103/PhysRevResearch.5.043251} {\bibfield  {journal} {\bibinfo  {journal} {Phys. Rev. Res.}\ }\textbf {\bibinfo {volume} {5}},\ \bibinfo {pages} {043251} (\bibinfo {year} {2023})}\BibitemShut {NoStop}%
\bibitem [{\citenamefont {Kawaguchi}\ and\ \citenamefont {Nakayama}(2013)}]{kawaguchi2013}%
  \BibitemOpen
  \bibfield  {author} {\bibinfo {author} {\bibfnamefont {K.}~\bibnamefont {Kawaguchi}}\ and\ \bibinfo {author} {\bibfnamefont {Y.}~\bibnamefont {Nakayama}},\ }\href {https://doi.org/10.1103/PhysRevE.88.022147} {\bibfield  {journal} {\bibinfo  {journal} {Phys. Rev. E}\ }\textbf {\bibinfo {volume} {88}},\ \bibinfo {pages} {022147} (\bibinfo {year} {2013})}\BibitemShut {NoStop}%
\bibitem [{\citenamefont {Shiraishi}\ and\ \citenamefont {Sagawa}(2015)}]{shiraishi2015}%
  \BibitemOpen
  \bibfield  {author} {\bibinfo {author} {\bibfnamefont {N.}~\bibnamefont {Shiraishi}}\ and\ \bibinfo {author} {\bibfnamefont {T.}~\bibnamefont {Sagawa}},\ }\href {https://doi.org/10.1103/PhysRevE.91.012130} {\bibfield  {journal} {\bibinfo  {journal} {Phys. Rev. E}\ }\textbf {\bibinfo {volume} {91}},\ \bibinfo {pages} {012130} (\bibinfo {year} {2015})}\BibitemShut {NoStop}%
\bibitem [{\citenamefont {Barato}\ and\ \citenamefont {Seifert}(2015{\natexlab{a}})}]{barato2015}%
  \BibitemOpen
  \bibfield  {author} {\bibinfo {author} {\bibfnamefont {A.~C.}\ \bibnamefont {Barato}}\ and\ \bibinfo {author} {\bibfnamefont {U.}~\bibnamefont {Seifert}},\ }\href {https://doi.org/10.1103/PhysRevLett.114.158101} {\bibfield  {journal} {\bibinfo  {journal} {Phys. Rev. Lett.}\ }\textbf {\bibinfo {volume} {114}},\ \bibinfo {pages} {158101} (\bibinfo {year} {2015}{\natexlab{a}})}\BibitemShut {NoStop}%
\bibitem [{\citenamefont {Gingrich}\ \emph {et~al.}(2016)\citenamefont {Gingrich}, \citenamefont {Horowitz}, \citenamefont {Perunov},\ and\ \citenamefont {England}}]{gingrich2016}%
  \BibitemOpen
  \bibfield  {author} {\bibinfo {author} {\bibfnamefont {T.~R.}\ \bibnamefont {Gingrich}}, \bibinfo {author} {\bibfnamefont {J.~M.}\ \bibnamefont {Horowitz}}, \bibinfo {author} {\bibfnamefont {N.}~\bibnamefont {Perunov}},\ and\ \bibinfo {author} {\bibfnamefont {J.~L.}\ \bibnamefont {England}},\ }\href {https://doi.org/10.1103/PhysRevLett.116.120601} {\bibfield  {journal} {\bibinfo  {journal} {Phys. Rev. Lett.}\ }\textbf {\bibinfo {volume} {116}},\ \bibinfo {pages} {120601} (\bibinfo {year} {2016})}\BibitemShut {NoStop}%
\bibitem [{\citenamefont {Pietzonka}\ \emph {et~al.}(2016)\citenamefont {Pietzonka}, \citenamefont {Barato},\ and\ \citenamefont {Seifert}}]{pietzonka2016}%
  \BibitemOpen
  \bibfield  {author} {\bibinfo {author} {\bibfnamefont {P.}~\bibnamefont {Pietzonka}}, \bibinfo {author} {\bibfnamefont {A.~C.}\ \bibnamefont {Barato}},\ and\ \bibinfo {author} {\bibfnamefont {U.}~\bibnamefont {Seifert}},\ }\href {https://doi.org/10.1103/PhysRevE.93.052145} {\bibfield  {journal} {\bibinfo  {journal} {Phys. Rev. E}\ }\textbf {\bibinfo {volume} {93}},\ \bibinfo {pages} {052145} (\bibinfo {year} {2016})}\BibitemShut {NoStop}%
\bibitem [{\citenamefont {Horowitz}\ and\ \citenamefont {Gingrich}(2020)}]{horowitz2020}%
  \BibitemOpen
  \bibfield  {author} {\bibinfo {author} {\bibfnamefont {J.~M.}\ \bibnamefont {Horowitz}}\ and\ \bibinfo {author} {\bibfnamefont {T.~R.}\ \bibnamefont {Gingrich}},\ }\href {https://doi.org/10.1038/s41567-019-0702-6} {\bibfield  {journal} {\bibinfo  {journal} {Nat. Phys.}\ }\textbf {\bibinfo {volume} {16}},\ \bibinfo {pages} {15} (\bibinfo {year} {2020})}\BibitemShut {NoStop}%
\bibitem [{\citenamefont {Rold\'an}\ and\ \citenamefont {Parrondo}(2010)}]{roldan2010}%
  \BibitemOpen
  \bibfield  {author} {\bibinfo {author} {\bibfnamefont {E.}~\bibnamefont {Rold\'an}}\ and\ \bibinfo {author} {\bibfnamefont {J.~M.~R.}\ \bibnamefont {Parrondo}},\ }\href {https://doi.org/10.1103/PhysRevLett.105.150607} {\bibfield  {journal} {\bibinfo  {journal} {Phys. Rev. Lett.}\ }\textbf {\bibinfo {volume} {105}},\ \bibinfo {pages} {150607} (\bibinfo {year} {2010})}\BibitemShut {NoStop}%
\bibitem [{\citenamefont {Rold\'an}\ and\ \citenamefont {Parrondo}(2012)}]{roldan2012}%
  \BibitemOpen
  \bibfield  {author} {\bibinfo {author} {\bibfnamefont {E.}~\bibnamefont {Rold\'an}}\ and\ \bibinfo {author} {\bibfnamefont {J.~M.~R.}\ \bibnamefont {Parrondo}},\ }\href {https://doi.org/10.1103/PhysRevE.85.031129} {\bibfield  {journal} {\bibinfo  {journal} {Phys. Rev. E}\ }\textbf {\bibinfo {volume} {85}},\ \bibinfo {pages} {031129} (\bibinfo {year} {2012})}\BibitemShut {NoStop}%
\bibitem [{\citenamefont {Biddle}\ and\ \citenamefont {Gunawardena}(2020)}]{biddle2020}%
  \BibitemOpen
  \bibfield  {author} {\bibinfo {author} {\bibfnamefont {J.~W.}\ \bibnamefont {Biddle}}\ and\ \bibinfo {author} {\bibfnamefont {J.}~\bibnamefont {Gunawardena}},\ }\href {https://doi.org/10.1103/PhysRevE.101.062125} {\bibfield  {journal} {\bibinfo  {journal} {Phys. Rev. E}\ }\textbf {\bibinfo {volume} {101}},\ \bibinfo {pages} {062125} (\bibinfo {year} {2020})}\BibitemShut {NoStop}%
\bibitem [{\citenamefont {Pietzonka}\ \emph {et~al.}(2021)\citenamefont {Pietzonka}, \citenamefont {Guioth},\ and\ \citenamefont {Jack}}]{pietzonka2021}%
  \BibitemOpen
  \bibfield  {author} {\bibinfo {author} {\bibfnamefont {P.}~\bibnamefont {Pietzonka}}, \bibinfo {author} {\bibfnamefont {J.}~\bibnamefont {Guioth}},\ and\ \bibinfo {author} {\bibfnamefont {R.~L.}\ \bibnamefont {Jack}},\ }\href {https://doi.org/10.1103/PhysRevE.104.064137} {\bibfield  {journal} {\bibinfo  {journal} {Phys. Rev. E}\ }\textbf {\bibinfo {volume} {104}},\ \bibinfo {pages} {064137} (\bibinfo {year} {2021})}\BibitemShut {NoStop}%
\bibitem [{\citenamefont {Pietzonka}\ and\ \citenamefont {Coghi}(2024)}]{pietzonka2023}%
  \BibitemOpen
  \bibfield  {author} {\bibinfo {author} {\bibfnamefont {P.}~\bibnamefont {Pietzonka}}\ and\ \bibinfo {author} {\bibfnamefont {F.}~\bibnamefont {Coghi}},\ }\href {https://doi.org/10.1103/PhysRevE.109.064128} {\bibfield  {journal} {\bibinfo  {journal} {Phys. Rev. E}\ }\textbf {\bibinfo {volume} {109}},\ \bibinfo {pages} {064128} (\bibinfo {year} {2024})}\BibitemShut {NoStop}%
\bibitem [{\citenamefont {Di~Terlizzi}\ \emph {et~al.}(2024{\natexlab{a}})\citenamefont {Di~Terlizzi}, \citenamefont {Gironella}, \citenamefont {Herraez-Aguilar}, \citenamefont {Betz}, \citenamefont {Monroy}, \citenamefont {Baiesi},\ and\ \citenamefont {Ritort}}]{Diterlizzi2024}%
  \BibitemOpen
  \bibfield  {author} {\bibinfo {author} {\bibfnamefont {I.}~\bibnamefont {Di~Terlizzi}}, \bibinfo {author} {\bibfnamefont {M.}~\bibnamefont {Gironella}}, \bibinfo {author} {\bibfnamefont {D.}~\bibnamefont {Herraez-Aguilar}}, \bibinfo {author} {\bibfnamefont {T.}~\bibnamefont {Betz}}, \bibinfo {author} {\bibfnamefont {F.}~\bibnamefont {Monroy}}, \bibinfo {author} {\bibfnamefont {M.}~\bibnamefont {Baiesi}},\ and\ \bibinfo {author} {\bibfnamefont {F.}~\bibnamefont {Ritort}},\ }\href {https://doi.org/10.1126/science.adh1823} {\bibfield  {journal} {\bibinfo  {journal} {Science}\ }\textbf {\bibinfo {volume} {383}},\ \bibinfo {pages} {971–976} (\bibinfo {year} {2024}{\natexlab{a}})}\BibitemShut {NoStop}%
\bibitem [{\citenamefont {Di~Terlizzi}\ \emph {et~al.}(2024{\natexlab{b}})\citenamefont {Di~Terlizzi}, \citenamefont {Baiesi},\ and\ \citenamefont {Ritort}}]{Diterlizzi2024_2}%
  \BibitemOpen
  \bibfield  {author} {\bibinfo {author} {\bibfnamefont {I.}~\bibnamefont {Di~Terlizzi}}, \bibinfo {author} {\bibfnamefont {M.}~\bibnamefont {Baiesi}},\ and\ \bibinfo {author} {\bibfnamefont {F.}~\bibnamefont {Ritort}},\ }\href {https://doi.org/10.1088/1367-2630/ad4fb9} {\bibfield  {journal} {\bibinfo  {journal} {New J. Phys.}\ }\textbf {\bibinfo {volume} {26}},\ \bibinfo {pages} {063013} (\bibinfo {year} {2024}{\natexlab{b}})}\BibitemShut {NoStop}%
\bibitem [{\citenamefont {Neri}\ \emph {et~al.}(2017)\citenamefont {Neri}, \citenamefont {Rold\'an},\ and\ \citenamefont {J\"ulicher}}]{neri2017}%
  \BibitemOpen
  \bibfield  {author} {\bibinfo {author} {\bibfnamefont {I.}~\bibnamefont {Neri}}, \bibinfo {author} {\bibfnamefont {E.}~\bibnamefont {Rold\'an}},\ and\ \bibinfo {author} {\bibfnamefont {F.}~\bibnamefont {J\"ulicher}},\ }\href {https://doi.org/10.1103/PhysRevX.7.011019} {\bibfield  {journal} {\bibinfo  {journal} {Phys. Rev. X}\ }\textbf {\bibinfo {volume} {7}},\ \bibinfo {pages} {011019} (\bibinfo {year} {2017})}\BibitemShut {NoStop}%
\bibitem [{\citenamefont {Neri}(2020)}]{neri2020}%
  \BibitemOpen
  \bibfield  {author} {\bibinfo {author} {\bibfnamefont {I.}~\bibnamefont {Neri}},\ }\href {https://doi.org/10.1103/PhysRevLett.124.040601} {\bibfield  {journal} {\bibinfo  {journal} {Phys. Rev. Lett.}\ }\textbf {\bibinfo {volume} {124}},\ \bibinfo {pages} {040601} (\bibinfo {year} {2020})}\BibitemShut {NoStop}%
\bibitem [{\citenamefont {Neri}(2022)}]{neri2022}%
  \BibitemOpen
  \bibfield  {author} {\bibinfo {author} {\bibfnamefont {I.}~\bibnamefont {Neri}},\ }\href {https://doi.org/10.1088/1751-8121/ac736b} {\bibfield  {journal} {\bibinfo  {journal} {J. Phys. A: Math. Theor.}\ }\textbf {\bibinfo {volume} {55}},\ \bibinfo {pages} {304005} (\bibinfo {year} {2022})}\BibitemShut {NoStop}%
\bibitem [{\citenamefont {Polettini}\ and\ \citenamefont {Neri}(2024)}]{polettini2024}%
  \BibitemOpen
  \bibfield  {author} {\bibinfo {author} {\bibfnamefont {M.}~\bibnamefont {Polettini}}\ and\ \bibinfo {author} {\bibfnamefont {I.}~\bibnamefont {Neri}},\ }\bibfield  {journal} {\bibinfo  {journal} {J. Stat. Phys.}\ }\textbf {\bibinfo {volume} {191}},\ \href {https://doi.org/10.1007/s10955-024-03236-5} {10.1007/s10955-024-03236-5} (\bibinfo {year} {2024})\BibitemShut {NoStop}%
\bibitem [{\citenamefont {Oberreiter}\ \emph {et~al.}(2022)\citenamefont {Oberreiter}, \citenamefont {Seifert},\ and\ \citenamefont {Barato}}]{oberreiter2022}%
  \BibitemOpen
  \bibfield  {author} {\bibinfo {author} {\bibfnamefont {L.}~\bibnamefont {Oberreiter}}, \bibinfo {author} {\bibfnamefont {U.}~\bibnamefont {Seifert}},\ and\ \bibinfo {author} {\bibfnamefont {A.~C.}\ \bibnamefont {Barato}},\ }\href {https://doi.org/10.1103/PhysRevE.106.014106} {\bibfield  {journal} {\bibinfo  {journal} {Phys. Rev. E}\ }\textbf {\bibinfo {volume} {106}},\ \bibinfo {pages} {014106} (\bibinfo {year} {2022})}\BibitemShut {NoStop}%
\bibitem [{\citenamefont {Ohga}\ \emph {et~al.}(2023)\citenamefont {Ohga}, \citenamefont {Ito},\ and\ \citenamefont {Kolchinsky}}]{ohga2023}%
  \BibitemOpen
  \bibfield  {author} {\bibinfo {author} {\bibfnamefont {N.}~\bibnamefont {Ohga}}, \bibinfo {author} {\bibfnamefont {S.}~\bibnamefont {Ito}},\ and\ \bibinfo {author} {\bibfnamefont {A.}~\bibnamefont {Kolchinsky}},\ }\href {https://doi.org/10.1103/PhysRevLett.131.077101} {\bibfield  {journal} {\bibinfo  {journal} {Phys. Rev. Lett.}\ }\textbf {\bibinfo {volume} {131}},\ \bibinfo {pages} {077101} (\bibinfo {year} {2023})}\BibitemShut {NoStop}%
\bibitem [{\citenamefont {Dechant}\ \emph {et~al.}(2023)\citenamefont {Dechant}, \citenamefont {Garnier-Brun},\ and\ \citenamefont {Sasa}}]{dechant2023}%
  \BibitemOpen
  \bibfield  {author} {\bibinfo {author} {\bibfnamefont {A.}~\bibnamefont {Dechant}}, \bibinfo {author} {\bibfnamefont {J.}~\bibnamefont {Garnier-Brun}},\ and\ \bibinfo {author} {\bibfnamefont {S.-i.}\ \bibnamefont {Sasa}},\ }\href {https://doi.org/10.1103/PhysRevLett.131.167101} {\bibfield  {journal} {\bibinfo  {journal} {Phys. Rev. Lett.}\ }\textbf {\bibinfo {volume} {131}},\ \bibinfo {pages} {167101} (\bibinfo {year} {2023})}\BibitemShut {NoStop}%
\bibitem [{\citenamefont {Berezhkovskii}\ and\ \citenamefont {Makarov}(2019)}]{berezhkovskii2019}%
  \BibitemOpen
  \bibfield  {author} {\bibinfo {author} {\bibfnamefont {A.~M.}\ \bibnamefont {Berezhkovskii}}\ and\ \bibinfo {author} {\bibfnamefont {D.~E.}\ \bibnamefont {Makarov}},\ }\href {https://doi.org/10.1063/1.5109293} {\bibfield  {journal} {\bibinfo  {journal} {J. Chem. Phys.}\ }\textbf {\bibinfo {volume} {151}},\ \bibinfo {pages} {065102} (\bibinfo {year} {2019})}\BibitemShut {NoStop}%
\bibitem [{\citenamefont {Martínez}\ \emph {et~al.}(2019)\citenamefont {Martínez}, \citenamefont {Bisker}, \citenamefont {Horowitz},\ and\ \citenamefont {Parrondo}}]{martinez2019}%
  \BibitemOpen
  \bibfield  {author} {\bibinfo {author} {\bibfnamefont {I.~A.}\ \bibnamefont {Martínez}}, \bibinfo {author} {\bibfnamefont {G.}~\bibnamefont {Bisker}}, \bibinfo {author} {\bibfnamefont {J.~M.}\ \bibnamefont {Horowitz}},\ and\ \bibinfo {author} {\bibfnamefont {J.~M.~R.}\ \bibnamefont {Parrondo}},\ }\href {https://doi.org/10.1038/s41467-019-11051-w} {\bibfield  {journal} {\bibinfo  {journal} {Nat. Commun.}\ }\textbf {\bibinfo {volume} {10}},\ \bibinfo {pages} {3542} (\bibinfo {year} {2019})}\BibitemShut {NoStop}%
\bibitem [{\citenamefont {Skinner}\ and\ \citenamefont {Dunkel}(2021{\natexlab{b}})}]{skinner2021_2}%
  \BibitemOpen
  \bibfield  {author} {\bibinfo {author} {\bibfnamefont {D.~J.}\ \bibnamefont {Skinner}}\ and\ \bibinfo {author} {\bibfnamefont {J.}~\bibnamefont {Dunkel}},\ }\href {https://doi.org/10.1103/PhysRevLett.127.198101} {\bibfield  {journal} {\bibinfo  {journal} {Phys. Rev. Lett.}\ }\textbf {\bibinfo {volume} {127}},\ \bibinfo {pages} {198101} (\bibinfo {year} {2021}{\natexlab{b}})}\BibitemShut {NoStop}%
\bibitem [{\citenamefont {Hartich}\ and\ \citenamefont {Godec}(2021)}]{hartich2021_2}%
  \BibitemOpen
  \bibfield  {author} {\bibinfo {author} {\bibfnamefont {D.}~\bibnamefont {Hartich}}\ and\ \bibinfo {author} {\bibfnamefont {A.}~\bibnamefont {Godec}},\ }\href {https://doi.org/10.1103/PhysRevX.11.041047} {\bibfield  {journal} {\bibinfo  {journal} {Phys. Rev. X}\ }\textbf {\bibinfo {volume} {11}},\ \bibinfo {pages} {041047} (\bibinfo {year} {2021})}\BibitemShut {NoStop}%
\bibitem [{\citenamefont {Hartich}\ and\ \citenamefont {Godec}(2023)}]{hartich2021}%
  \BibitemOpen
  \bibfield  {author} {\bibinfo {author} {\bibfnamefont {D.}~\bibnamefont {Hartich}}\ and\ \bibinfo {author} {\bibfnamefont {A.}~\bibnamefont {Godec}},\ }\href {https://doi.org/10.1103/PhysRevResearch.5.L032017} {\bibfield  {journal} {\bibinfo  {journal} {Phys. Rev. Res.}\ }\textbf {\bibinfo {volume} {5}},\ \bibinfo {pages} {L032017} (\bibinfo {year} {2023})}\BibitemShut {NoStop}%
\bibitem [{\citenamefont {Wang}\ and\ \citenamefont {Qian}(2007)}]{qian2007}%
  \BibitemOpen
  \bibfield  {author} {\bibinfo {author} {\bibfnamefont {H.}~\bibnamefont {Wang}}\ and\ \bibinfo {author} {\bibfnamefont {H.}~\bibnamefont {Qian}},\ }\href {https://doi.org/10.1063/1.2432065} {\bibfield  {journal} {\bibinfo  {journal} {J. Math. Phys.}\ }\textbf {\bibinfo {volume} {48}},\ \bibinfo {pages} {013303} (\bibinfo {year} {2007})}\BibitemShut {NoStop}%
\bibitem [{\citenamefont {Esposito}\ and\ \citenamefont {Lindenberg}(2008)}]{esposito2008}%
  \BibitemOpen
  \bibfield  {author} {\bibinfo {author} {\bibfnamefont {M.}~\bibnamefont {Esposito}}\ and\ \bibinfo {author} {\bibfnamefont {K.}~\bibnamefont {Lindenberg}},\ }\href {https://doi.org/10.1103/PhysRevE.77.051119} {\bibfield  {journal} {\bibinfo  {journal} {Phys. Rev. E}\ }\textbf {\bibinfo {volume} {77}},\ \bibinfo {pages} {051119} (\bibinfo {year} {2008})}\BibitemShut {NoStop}%
\bibitem [{\citenamefont {Maes}\ \emph {et~al.}(2009)\citenamefont {Maes}, \citenamefont {Neto{\v{c}}n{\'{y}}},\ and\ \citenamefont {Wynants}}]{maes2009}%
  \BibitemOpen
  \bibfield  {author} {\bibinfo {author} {\bibfnamefont {C.}~\bibnamefont {Maes}}, \bibinfo {author} {\bibfnamefont {K.}~\bibnamefont {Neto{\v{c}}n{\'{y}}}},\ and\ \bibinfo {author} {\bibfnamefont {B.}~\bibnamefont {Wynants}},\ }\href {https://doi.org/10.1088/1751-8113/42/36/365002} {\bibfield  {journal} {\bibinfo  {journal} {J. Phys. A: Math. Theor.}\ }\textbf {\bibinfo {volume} {42}},\ \bibinfo {pages} {365002} (\bibinfo {year} {2009})}\BibitemShut {NoStop}%
\bibitem [{\citenamefont {Ertel}\ \emph {et~al.}(2022)\citenamefont {Ertel}, \citenamefont {van~der Meer},\ and\ \citenamefont {Seifert}}]{PRE}%
  \BibitemOpen
  \bibfield  {author} {\bibinfo {author} {\bibfnamefont {B.}~\bibnamefont {Ertel}}, \bibinfo {author} {\bibfnamefont {J.}~\bibnamefont {van~der Meer}},\ and\ \bibinfo {author} {\bibfnamefont {U.}~\bibnamefont {Seifert}},\ }\href {https://doi.org/10.1103/PhysRevE.105.044113} {\bibfield  {journal} {\bibinfo  {journal} {Phys. Rev. E}\ }\textbf {\bibinfo {volume} {105}},\ \bibinfo {pages} {044113} (\bibinfo {year} {2022})}\BibitemShut {NoStop}%
\bibitem [{\citenamefont {Hartich}\ and\ \citenamefont {Godec}(2024)}]{hartich2021_comment}%
  \BibitemOpen
  \bibfield  {author} {\bibinfo {author} {\bibfnamefont {D.}~\bibnamefont {Hartich}}\ and\ \bibinfo {author} {\bibfnamefont {A.}~\bibnamefont {Godec}},\ }\href {https://doi.org/10.1038/s41467-024-52602-0} {\bibfield  {journal} {\bibinfo  {journal} {Nat. Commun.}\ }\textbf {\bibinfo {volume} {15}},\ \bibinfo {pages} {8678} (\bibinfo {year} {2024})}\BibitemShut {NoStop}%
\bibitem [{\citenamefont {Bisker}\ \emph {et~al.}(2024)\citenamefont {Bisker}, \citenamefont {Martínez}, \citenamefont {Horowitz},\ and\ \citenamefont {Parrondo}}]{bisker2022_Reply}%
  \BibitemOpen
  \bibfield  {author} {\bibinfo {author} {\bibfnamefont {G.}~\bibnamefont {Bisker}}, \bibinfo {author} {\bibfnamefont {I.~A.}\ \bibnamefont {Martínez}}, \bibinfo {author} {\bibfnamefont {J.~M.}\ \bibnamefont {Horowitz}},\ and\ \bibinfo {author} {\bibfnamefont {J.~M.~R.}\ \bibnamefont {Parrondo}},\ }\href {https://doi.org/10.1038/s41467-024-52603-z} {\bibfield  {journal} {\bibinfo  {journal} {Nat. Commun.}\ }\textbf {\bibinfo {volume} {15}},\ \bibinfo {pages} {8679} (\bibinfo {year} {2024})}\BibitemShut {NoStop}%
\bibitem [{\citenamefont {van~der Meer}\ \emph {et~al.}(2023)\citenamefont {van~der Meer}, \citenamefont {Deg\"unther},\ and\ \citenamefont {Seifert}}]{PRL}%
  \BibitemOpen
  \bibfield  {author} {\bibinfo {author} {\bibfnamefont {J.}~\bibnamefont {van~der Meer}}, \bibinfo {author} {\bibfnamefont {J.}~\bibnamefont {Deg\"unther}},\ and\ \bibinfo {author} {\bibfnamefont {U.}~\bibnamefont {Seifert}},\ }\href {https://doi.org/10.1103/PhysRevLett.130.257101} {\bibfield  {journal} {\bibinfo  {journal} {Phys. Rev. Lett.}\ }\textbf {\bibinfo {volume} {130}},\ \bibinfo {pages} {257101} (\bibinfo {year} {2023})}\BibitemShut {NoStop}%
\bibitem [{\citenamefont {Deg\"{u}nther}\ \emph {et~al.}(2024{\natexlab{a}})\citenamefont {Deg\"{u}nther}, \citenamefont {van~der Meer},\ and\ \citenamefont {Seifert}}]{PRR}%
  \BibitemOpen
  \bibfield  {author} {\bibinfo {author} {\bibfnamefont {J.}~\bibnamefont {Deg\"{u}nther}}, \bibinfo {author} {\bibfnamefont {J.}~\bibnamefont {van~der Meer}},\ and\ \bibinfo {author} {\bibfnamefont {U.}~\bibnamefont {Seifert}},\ }\href {https://doi.org/10.1103/PhysRevResearch.6.023175} {\bibfield  {journal} {\bibinfo  {journal} {Phys. Rev. Res.}\ }\textbf {\bibinfo {volume} {6}},\ \bibinfo {pages} {023175} (\bibinfo {year} {2024}{\natexlab{a}})}\BibitemShut {NoStop}%
\bibitem [{\citenamefont {Deg\"{u}nther}\ \emph {et~al.}(2024{\natexlab{b}})\citenamefont {Deg\"{u}nther}, \citenamefont {van~der Meer},\ and\ \citenamefont {Seifert}}]{PNAS}%
  \BibitemOpen
  \bibfield  {author} {\bibinfo {author} {\bibfnamefont {J.}~\bibnamefont {Deg\"{u}nther}}, \bibinfo {author} {\bibfnamefont {J.}~\bibnamefont {van~der Meer}},\ and\ \bibinfo {author} {\bibfnamefont {U.}~\bibnamefont {Seifert}},\ }\href {https://doi.org/10.1073/pnas.2405371121} {\bibfield  {journal} {\bibinfo  {journal} {Proc. Natl. Acad. Sci.}\ }\textbf {\bibinfo {volume} {121}},\ \bibinfo {pages} {e2405371121} (\bibinfo {year} {2024}{\natexlab{b}})}\BibitemShut {NoStop}%
\bibitem [{\citenamefont {van~der Meer}\ \emph {et~al.}(2022)\citenamefont {van~der Meer}, \citenamefont {Ertel},\ and\ \citenamefont {Seifert}}]{PRX}%
  \BibitemOpen
  \bibfield  {author} {\bibinfo {author} {\bibfnamefont {J.}~\bibnamefont {van~der Meer}}, \bibinfo {author} {\bibfnamefont {B.}~\bibnamefont {Ertel}},\ and\ \bibinfo {author} {\bibfnamefont {U.}~\bibnamefont {Seifert}},\ }\href {https://doi.org/10.1103/PhysRevX.12.031025} {\bibfield  {journal} {\bibinfo  {journal} {Phys. Rev. X}\ }\textbf {\bibinfo {volume} {12}},\ \bibinfo {pages} {031025} (\bibinfo {year} {2022})}\BibitemShut {NoStop}%
\bibitem [{\citenamefont {Harunari}\ \emph {et~al.}(2022)\citenamefont {Harunari}, \citenamefont {Dutta}, \citenamefont {Polettini},\ and\ \citenamefont {Rold\'an}}]{harunari2022}%
  \BibitemOpen
  \bibfield  {author} {\bibinfo {author} {\bibfnamefont {P.~E.}\ \bibnamefont {Harunari}}, \bibinfo {author} {\bibfnamefont {A.}~\bibnamefont {Dutta}}, \bibinfo {author} {\bibfnamefont {M.}~\bibnamefont {Polettini}},\ and\ \bibinfo {author} {\bibfnamefont {E.}~\bibnamefont {Rold\'an}},\ }\href {https://doi.org/10.1103/PhysRevX.12.041026} {\bibfield  {journal} {\bibinfo  {journal} {Phys. Rev. X}\ }\textbf {\bibinfo {volume} {12}},\ \bibinfo {pages} {041026} (\bibinfo {year} {2022})}\BibitemShut {NoStop}%
\bibitem [{\citenamefont {Cao}\ and\ \citenamefont {Silbey}(2008)}]{cao2008}%
  \BibitemOpen
  \bibfield  {author} {\bibinfo {author} {\bibfnamefont {J.}~\bibnamefont {Cao}}\ and\ \bibinfo {author} {\bibfnamefont {R.~J.}\ \bibnamefont {Silbey}},\ }\href {https://doi.org/10.1021/jp803347m} {\bibfield  {journal} {\bibinfo  {journal} {J. Phys. Chem. B}\ }\textbf {\bibinfo {volume} {112}},\ \bibinfo {pages} {12867–12880} (\bibinfo {year} {2008})}\BibitemShut {NoStop}%
\bibitem [{\citenamefont {Li}\ and\ \citenamefont {Kolomeisky}(2013)}]{li2013}%
  \BibitemOpen
  \bibfield  {author} {\bibinfo {author} {\bibfnamefont {X.}~\bibnamefont {Li}}\ and\ \bibinfo {author} {\bibfnamefont {A.~B.}\ \bibnamefont {Kolomeisky}},\ }\bibfield  {journal} {\bibinfo  {journal} {J. Chem. Phys}\ }\textbf {\bibinfo {volume} {139}},\ \href {https://doi.org/10.1063/1.4824392} {10.1063/1.4824392} (\bibinfo {year} {2013})\BibitemShut {NoStop}%
\bibitem [{\citenamefont {Panigrahy}\ \emph {et~al.}(2019)\citenamefont {Panigrahy}, \citenamefont {Kumar}, \citenamefont {Chowdhury},\ and\ \citenamefont {Dua}}]{panigrahy2019}%
  \BibitemOpen
  \bibfield  {author} {\bibinfo {author} {\bibfnamefont {M.}~\bibnamefont {Panigrahy}}, \bibinfo {author} {\bibfnamefont {A.}~\bibnamefont {Kumar}}, \bibinfo {author} {\bibfnamefont {S.}~\bibnamefont {Chowdhury}},\ and\ \bibinfo {author} {\bibfnamefont {A.}~\bibnamefont {Dua}},\ }\bibfield  {journal} {\bibinfo  {journal} {J. Chem. Phys.}\ }\textbf {\bibinfo {volume} {150}},\ \href {https://doi.org/10.1063/1.5087974} {10.1063/1.5087974} (\bibinfo {year} {2019})\BibitemShut {NoStop}%
\bibitem [{\citenamefont {Thorneywork}\ \emph {et~al.}(2020)\citenamefont {Thorneywork}, \citenamefont {Gladrow}, \citenamefont {Qing}, \citenamefont {Rico-Pasto}, \citenamefont {Ritort}, \citenamefont {Bayley}, \citenamefont {Kolomeisky},\ and\ \citenamefont {Keyser}}]{thorneywork2020}%
  \BibitemOpen
  \bibfield  {author} {\bibinfo {author} {\bibfnamefont {A.~L.}\ \bibnamefont {Thorneywork}}, \bibinfo {author} {\bibfnamefont {J.}~\bibnamefont {Gladrow}}, \bibinfo {author} {\bibfnamefont {Y.}~\bibnamefont {Qing}}, \bibinfo {author} {\bibfnamefont {M.}~\bibnamefont {Rico-Pasto}}, \bibinfo {author} {\bibfnamefont {F.}~\bibnamefont {Ritort}}, \bibinfo {author} {\bibfnamefont {H.}~\bibnamefont {Bayley}}, \bibinfo {author} {\bibfnamefont {A.~B.}\ \bibnamefont {Kolomeisky}},\ and\ \bibinfo {author} {\bibfnamefont {U.~F.}\ \bibnamefont {Keyser}},\ }\href {https://doi.org/10.1126/sciadv.aaz4642} {\bibfield  {journal} {\bibinfo  {journal} {Sci. Adv.}\ }\textbf {\bibinfo {volume} {6}},\ \bibinfo {pages} {eaaz4642} (\bibinfo {year} {2020})}\BibitemShut {NoStop}%
\bibitem [{\citenamefont {Berezhkovskii}\ and\ \citenamefont {Makarov}(2020)}]{berezhkovskii2020}%
  \BibitemOpen
  \bibfield  {author} {\bibinfo {author} {\bibfnamefont {A.~M.}\ \bibnamefont {Berezhkovskii}}\ and\ \bibinfo {author} {\bibfnamefont {D.~E.}\ \bibnamefont {Makarov}},\ }\href {https://doi.org/10.1021/acs.jpclett.9b03705} {\bibfield  {journal} {\bibinfo  {journal} {J. Phys. Chem. Lett.}\ }\textbf {\bibinfo {volume} {11}},\ \bibinfo {pages} {1682} (\bibinfo {year} {2020})}\BibitemShut {NoStop}%
\bibitem [{\citenamefont {Satija}\ \emph {et~al.}(2020)\citenamefont {Satija}, \citenamefont {Berezhkovskii},\ and\ \citenamefont {Makarov}}]{satija2020}%
  \BibitemOpen
  \bibfield  {author} {\bibinfo {author} {\bibfnamefont {R.}~\bibnamefont {Satija}}, \bibinfo {author} {\bibfnamefont {A.~M.}\ \bibnamefont {Berezhkovskii}},\ and\ \bibinfo {author} {\bibfnamefont {D.~E.}\ \bibnamefont {Makarov}},\ }\href {https://doi.org/10.1073/pnas.2008307117} {\bibfield  {journal} {\bibinfo  {journal} {Proc. Natl. Acad. Sci.}\ }\textbf {\bibinfo {volume} {117}},\ \bibinfo {pages} {27116–27123} (\bibinfo {year} {2020})}\BibitemShut {NoStop}%
\bibitem [{\citenamefont {Ertel}\ \emph {et~al.}(2023)\citenamefont {Ertel}, \citenamefont {van~der Meer},\ and\ \citenamefont {Seifert}}]{IJMS}%
  \BibitemOpen
  \bibfield  {author} {\bibinfo {author} {\bibfnamefont {B.}~\bibnamefont {Ertel}}, \bibinfo {author} {\bibfnamefont {J.}~\bibnamefont {van~der Meer}},\ and\ \bibinfo {author} {\bibfnamefont {U.}~\bibnamefont {Seifert}},\ }\href {https://doi.org/10.3390/ijms24087610} {\bibfield  {journal} {\bibinfo  {journal} {Int. J. Mol. Sci.}\ }\textbf {\bibinfo {volume} {24}},\ \bibinfo {pages} {7610} (\bibinfo {year} {2023})}\BibitemShut {NoStop}%
\bibitem [{\citenamefont {Fersht}(2002)}]{fersht2002}%
  \BibitemOpen
  \bibfield  {author} {\bibinfo {author} {\bibfnamefont {A.}~\bibnamefont {Fersht}},\ }\href@noop {} {\emph {\bibinfo {title} {Structure and mechanism in protein science: A Guide to Enzyme Catalysis and Protein Folding}}},\ \bibinfo {edition} {4th}\ ed.\ (\bibinfo  {publisher} {W.H. Freeman},\ \bibinfo {address} {New York, NY},\ \bibinfo {year} {2002})\BibitemShut {NoStop}%
\bibitem [{\citenamefont {Illanes}(2008)}]{Illanes2008}%
  \BibitemOpen
  \bibinfo {editor} {\bibfnamefont {A.}~\bibnamefont {Illanes}},\ ed.,\ \href@noop {} {\emph {\bibinfo {title} {Enzyme Biocatalysis-Principles and Applications}}}\ (\bibinfo  {publisher} {Springer Dordrecht},\ \bibinfo {address} {New York, NY},\ \bibinfo {year} {2008})\BibitemShut {NoStop}%
\bibitem [{\citenamefont {Cornish-Bowden}(2012)}]{cornish2012}%
  \BibitemOpen
  \bibfield  {author} {\bibinfo {author} {\bibfnamefont {A.}~\bibnamefont {Cornish-Bowden}},\ }\href@noop {} {\emph {\bibinfo {title} {Fundamentals of enzyme kinetics}}},\ \bibinfo {edition} {4th}\ ed.\ (\bibinfo  {publisher} {Wiley-VCH Verlag},\ \bibinfo {address} {Weinheim, Germany},\ \bibinfo {year} {2012})\BibitemShut {NoStop}%
\bibitem [{\citenamefont {Godec}\ and\ \citenamefont {Makarov}(2022)}]{godec2022}%
  \BibitemOpen
  \bibfield  {author} {\bibinfo {author} {\bibfnamefont {A.}~\bibnamefont {Godec}}\ and\ \bibinfo {author} {\bibfnamefont {D.~E.}\ \bibnamefont {Makarov}},\ }\href {https://doi.org/10.1021/acs.jpclett.2c03244} {\bibfield  {journal} {\bibinfo  {journal} {J. Phys. Chem. Lett.}\ }\textbf {\bibinfo {volume} {14}},\ \bibinfo {pages} {49–56} (\bibinfo {year} {2022})}\BibitemShut {NoStop}%
\bibitem [{\citenamefont {Baiesi}\ \emph {et~al.}(2024)\citenamefont {Baiesi}, \citenamefont {Nishiyama},\ and\ \citenamefont {Falasco}}]{baiesi2024}%
  \BibitemOpen
  \bibfield  {author} {\bibinfo {author} {\bibfnamefont {M.}~\bibnamefont {Baiesi}}, \bibinfo {author} {\bibfnamefont {T.}~\bibnamefont {Nishiyama}},\ and\ \bibinfo {author} {\bibfnamefont {G.}~\bibnamefont {Falasco}},\ }\href@noop {} {\bibfield  {journal} {\bibinfo  {journal} {Commun. Phys.}\ }\textbf {\bibinfo {volume} {7}} (\bibinfo {year} {2024})}\BibitemShut {NoStop}%
\bibitem [{\citenamefont {Blom}\ \emph {et~al.}(2024)\citenamefont {Blom}, \citenamefont {Song}, \citenamefont {Vouga}, \citenamefont {Godec},\ and\ \citenamefont {Makarov}}]{blom2024}%
  \BibitemOpen
  \bibfield  {author} {\bibinfo {author} {\bibfnamefont {K.}~\bibnamefont {Blom}}, \bibinfo {author} {\bibfnamefont {K.}~\bibnamefont {Song}}, \bibinfo {author} {\bibfnamefont {E.}~\bibnamefont {Vouga}}, \bibinfo {author} {\bibfnamefont {A.}~\bibnamefont {Godec}},\ and\ \bibinfo {author} {\bibfnamefont {D.~E.}\ \bibnamefont {Makarov}},\ }\href {https://doi.org/10.1073/pnas.2318333121} {\bibfield  {journal} {\bibinfo  {journal} {Proc. Natl. Acad. Sci.}\ }\textbf {\bibinfo {volume} {121}},\ \bibinfo {pages} {e2318333121} (\bibinfo {year} {2024})}\BibitemShut {NoStop}%
\bibitem [{\citenamefont {Song}\ \emph {et~al.}(2024)\citenamefont {Song}, \citenamefont {Makarov},\ and\ \citenamefont {Vouga}}]{song2024}%
  \BibitemOpen
  \bibfield  {author} {\bibinfo {author} {\bibfnamefont {K.}~\bibnamefont {Song}}, \bibinfo {author} {\bibfnamefont {D.~E.}\ \bibnamefont {Makarov}},\ and\ \bibinfo {author} {\bibfnamefont {E.}~\bibnamefont {Vouga}},\ }\bibfield  {journal} {\bibinfo  {journal} {J. Chem. Phys.}\ }\textbf {\bibinfo {volume} {161}},\ \href {https://doi.org/10.1063/5.0218040} {10.1063/5.0218040} (\bibinfo {year} {2024})\BibitemShut {NoStop}%
\bibitem [{\citenamefont {Ertel}\ and\ \citenamefont {Seifert}(2024)}]{PRE2}%
  \BibitemOpen
  \bibfield  {author} {\bibinfo {author} {\bibfnamefont {B.}~\bibnamefont {Ertel}}\ and\ \bibinfo {author} {\bibfnamefont {U.}~\bibnamefont {Seifert}},\ }\href {https://doi.org/10.1103/PhysRevE.109.054109} {\bibfield  {journal} {\bibinfo  {journal} {Phys. Rev. E}\ }\textbf {\bibinfo {volume} {109}},\ \bibinfo {pages} {054109} (\bibinfo {year} {2024})}\BibitemShut {NoStop}%
\bibitem [{\citenamefont {Harunari}(2024)}]{harunari2024}%
  \BibitemOpen
  \bibfield  {author} {\bibinfo {author} {\bibfnamefont {P.~E.}\ \bibnamefont {Harunari}},\ }\href {https://doi.org/10.1103/PhysRevE.110.024122} {\bibfield  {journal} {\bibinfo  {journal} {Phys. Rev. E}\ }\textbf {\bibinfo {volume} {110}},\ \bibinfo {pages} {024122} (\bibinfo {year} {2024})}\BibitemShut {NoStop}%
\bibitem [{\citenamefont {Bebon}\ and\ \citenamefont {Godec}(2023)}]{bebon2023}%
  \BibitemOpen
  \bibfield  {author} {\bibinfo {author} {\bibfnamefont {R.}~\bibnamefont {Bebon}}\ and\ \bibinfo {author} {\bibfnamefont {A.}~\bibnamefont {Godec}},\ }\href {https://doi.org/10.1103/PhysRevLett.131.237101} {\bibfield  {journal} {\bibinfo  {journal} {Phys. Rev. Lett.}\ }\textbf {\bibinfo {volume} {131}},\ \bibinfo {pages} {237101} (\bibinfo {year} {2023})}\BibitemShut {NoStop}%
\bibitem [{\citenamefont {Hill}(1989)}]{hill1989}%
  \BibitemOpen
  \bibfield  {author} {\bibinfo {author} {\bibfnamefont {T.~L.}\ \bibnamefont {Hill}},\ }\href {https://doi.org/10.1007/978-1-4612-3558-3} {\emph {\bibinfo {title} {Free Energy Transduction and Biochemical Cycle Kinetics}}}\ (\bibinfo  {publisher} {Springer New York},\ \bibinfo {year} {1989})\BibitemShut {NoStop}%
\bibitem [{\citenamefont {Jiang}\ \emph {et~al.}(2004)\citenamefont {Jiang}, \citenamefont {Qian},\ and\ \citenamefont {Qian}}]{jiang2004}%
  \BibitemOpen
  \bibfield  {author} {\bibinfo {author} {\bibfnamefont {D.-Q.}\ \bibnamefont {Jiang}}, \bibinfo {author} {\bibfnamefont {M.}~\bibnamefont {Qian}},\ and\ \bibinfo {author} {\bibfnamefont {M.-P.}\ \bibnamefont {Qian}},\ }\href {https://doi.org/10.1007/b94615} {\emph {\bibinfo {title} {Mathematical Theory of Nonequilibrium Steady States}}}\ (\bibinfo  {publisher} {Springer Berlin Heidelberg},\ \bibinfo {year} {2004})\BibitemShut {NoStop}%
\bibitem [{\citenamefont {Schnakenberg}(1976)}]{schnakenberg1976}%
  \BibitemOpen
  \bibfield  {author} {\bibinfo {author} {\bibfnamefont {J.}~\bibnamefont {Schnakenberg}},\ }\href {https://doi.org/10.1103/RevModPhys.48.571} {\bibfield  {journal} {\bibinfo  {journal} {Rev. Mod. Phys.}\ }\textbf {\bibinfo {volume} {48}},\ \bibinfo {pages} {571} (\bibinfo {year} {1976})}\BibitemShut {NoStop}%
\bibitem [{\citenamefont {Sekimoto}(2021)}]{sekimoto2021}%
  \BibitemOpen
  \bibfield  {author} {\bibinfo {author} {\bibfnamefont {K.}~\bibnamefont {Sekimoto}},\ }\href {https://arxiv.org/abs/2110.02216} {\bibfield  {journal} {\bibinfo  {journal} {arXiv:2110.02216 [cond-mat.stat-mech]}\ } (\bibinfo {year} {2021})}\BibitemShut {NoStop}%
\bibitem [{\citenamefont {Dechant}(2018)}]{dechant2018}%
  \BibitemOpen
  \bibfield  {author} {\bibinfo {author} {\bibfnamefont {A.}~\bibnamefont {Dechant}},\ }\href {https://doi.org/10.1088/1751-8121/aaf3ff} {\bibfield  {journal} {\bibinfo  {journal} {J. Phys. A: Math. Theor.}\ }\textbf {\bibinfo {volume} {52}},\ \bibinfo {pages} {035001} (\bibinfo {year} {2018})}\BibitemShut {NoStop}%
\bibitem [{\citenamefont {Hasegawa}\ and\ \citenamefont {Van~Vu}(2019)}]{hasegawa2019}%
  \BibitemOpen
  \bibfield  {author} {\bibinfo {author} {\bibfnamefont {Y.}~\bibnamefont {Hasegawa}}\ and\ \bibinfo {author} {\bibfnamefont {T.}~\bibnamefont {Van~Vu}},\ }\href {https://doi.org/10.1103/PhysRevE.99.062126} {\bibfield  {journal} {\bibinfo  {journal} {Phys. Rev. E}\ }\textbf {\bibinfo {volume} {99}},\ \bibinfo {pages} {062126} (\bibinfo {year} {2019})}\BibitemShut {NoStop}%
\bibitem [{\citenamefont {Dechant}\ and\ \citenamefont {Sasa}(2020)}]{dechant2020}%
  \BibitemOpen
  \bibfield  {author} {\bibinfo {author} {\bibfnamefont {A.}~\bibnamefont {Dechant}}\ and\ \bibinfo {author} {\bibfnamefont {S.-i.}\ \bibnamefont {Sasa}},\ }\href {https://doi.org/10.1073/pnas.1918386117} {\bibfield  {journal} {\bibinfo  {journal} {Proc. Natl. Acad. Sci.}\ }\textbf {\bibinfo {volume} {117}},\ \bibinfo {pages} {6430–6436} (\bibinfo {year} {2020})}\BibitemShut {NoStop}%
\bibitem [{\citenamefont {Cover}\ and\ \citenamefont {Thomas}(2006)}]{cover2006}%
  \BibitemOpen
  \bibfield  {author} {\bibinfo {author} {\bibfnamefont {T.~M.}\ \bibnamefont {Cover}}\ and\ \bibinfo {author} {\bibfnamefont {J.~A.}\ \bibnamefont {Thomas}},\ }\href@noop {} {\emph {\bibinfo {title} {Elements of Information Theory (Wiley Series in Telecommunications and Signal Processing)}}}\ (\bibinfo  {publisher} {Wiley-Interscience},\ \bibinfo {address} {USA},\ \bibinfo {year} {2006})\BibitemShut {NoStop}%
\bibitem [{\citenamefont {Gillespie}(1977)}]{gillespie1977}%
  \BibitemOpen
  \bibfield  {author} {\bibinfo {author} {\bibfnamefont {D.~T.}\ \bibnamefont {Gillespie}},\ }\href {https://doi.org/10.1021/j100540a008} {\bibfield  {journal} {\bibinfo  {journal} {J. Phys. Chem.}\ }\textbf {\bibinfo {volume} {81}},\ \bibinfo {pages} {2340} (\bibinfo {year} {1977})}\BibitemShut {NoStop}%
\bibitem [{\citenamefont {Lucente}\ \emph {et~al.}(2023)\citenamefont {Lucente}, \citenamefont {Puglisi}, \citenamefont {Viale},\ and\ \citenamefont {Vulpiani}}]{luce23}%
  \BibitemOpen
  \bibfield  {author} {\bibinfo {author} {\bibfnamefont {D.}~\bibnamefont {Lucente}}, \bibinfo {author} {\bibfnamefont {A.}~\bibnamefont {Puglisi}}, \bibinfo {author} {\bibfnamefont {M.}~\bibnamefont {Viale}},\ and\ \bibinfo {author} {\bibfnamefont {A.}~\bibnamefont {Vulpiani}},\ }\href {https://doi.org/10.1088/1742-5468/ad063b} {\bibfield  {journal} {\bibinfo  {journal} {Journal of Statistical Mechanics: Theory and Experiment}\ }\textbf {\bibinfo {volume} {2023}},\ \bibinfo {pages} {113202} (\bibinfo {year} {2023})}\BibitemShut {NoStop}%
\bibitem [{\citenamefont {Yu}\ and\ \citenamefont {Harunari}(2024)}]{harunari2024_2}%
  \BibitemOpen
  \bibfield  {author} {\bibinfo {author} {\bibfnamefont {Q.}~\bibnamefont {Yu}}\ and\ \bibinfo {author} {\bibfnamefont {P.~E.}\ \bibnamefont {Harunari}},\ }\href {https://doi.org/10.1088/1742-5468/ad8152} {\bibfield  {journal} {\bibinfo  {journal} {J. Stat. Mech.}\ }\textbf {\bibinfo {volume} {2024}},\ \bibinfo {pages} {103201} (\bibinfo {year} {2024})}\BibitemShut {NoStop}%
\bibitem [{\citenamefont {Lervik}\ \emph {et~al.}(2015)\citenamefont {Lervik}, \citenamefont {Kjelstrup},\ and\ \citenamefont {Qian}}]{lervik2015}%
  \BibitemOpen
  \bibfield  {author} {\bibinfo {author} {\bibfnamefont {A.}~\bibnamefont {Lervik}}, \bibinfo {author} {\bibfnamefont {S.}~\bibnamefont {Kjelstrup}},\ and\ \bibinfo {author} {\bibfnamefont {H.}~\bibnamefont {Qian}},\ }\href {https://doi.org/10.1039/c4cp04334k} {\bibfield  {journal} {\bibinfo  {journal} {Phys. Chem. Chem. Phys.}\ }\textbf {\bibinfo {volume} {17}},\ \bibinfo {pages} {1317–1324} (\bibinfo {year} {2015})}\BibitemShut {NoStop}%
\bibitem [{\citenamefont {Wachtel}\ \emph {et~al.}(2018)\citenamefont {Wachtel}, \citenamefont {Rao},\ and\ \citenamefont {Esposito}}]{wachtel2018}%
  \BibitemOpen
  \bibfield  {author} {\bibinfo {author} {\bibfnamefont {A.}~\bibnamefont {Wachtel}}, \bibinfo {author} {\bibfnamefont {R.}~\bibnamefont {Rao}},\ and\ \bibinfo {author} {\bibfnamefont {M.}~\bibnamefont {Esposito}},\ }\href@noop {} {\bibfield  {journal} {\bibinfo  {journal} {New J. Phys.}\ }\textbf {\bibinfo {volume} {20}},\ \bibinfo {pages} {042002} (\bibinfo {year} {2018})}\BibitemShut {NoStop}%
\bibitem [{\citenamefont {Barato}\ and\ \citenamefont {Seifert}(2015{\natexlab{b}})}]{barato2015_2}%
  \BibitemOpen
  \bibfield  {author} {\bibinfo {author} {\bibfnamefont {A.~C.}\ \bibnamefont {Barato}}\ and\ \bibinfo {author} {\bibfnamefont {U.}~\bibnamefont {Seifert}},\ }\href {https://doi.org/10.1021/acs.jpcb.5b01918} {\bibfield  {journal} {\bibinfo  {journal} {J. Phys. Chem. B}\ }\textbf {\bibinfo {volume} {119}},\ \bibinfo {pages} {6555–6561} (\bibinfo {year} {2015}{\natexlab{b}})}\BibitemShut {NoStop}%
\end{thebibliography}

\end{document}